\documentclass[10pt,twocolumn,aps,pre,superscriptaddress]{revtex4} %

\usepackage{amsmath}
\usepackage{graphicx}
\usepackage{color}
\usepackage{upgreek}
\usepackage[linkcolor = blue, citecolor = blue, urlcolor = blue, colorlinks = true]{hyperref}
\usepackage{amssymb}
\usepackage[version=3]{mhchem}
\usepackage{upgreek}
\usepackage[exponent-product = \times]{siunitx}
\DeclareSIUnit\eq{\text{eq}}
\DeclareSIUnit\Hertz{\text{Hz}}
\begin{document}

\author{Ran Niu}
\affiliation{Institut f{\"u}r Physik, Johannes Gutenberg Universit{\"a}t Mainz, 55128 Mainz, Germany}

\author{Patrick Kreissl}
\affiliation{Institute for Computational Physics, Universit{\"a}t Stuttgart, Allmandring 3, 70569 Stuttgart, Germany}

\author{Aidan T. Brown}
\affiliation{SUPA, School of Physics and Astronomy, The University of Edinburgh, King's Buildings, Peter Guthrie Tait Road, Edinburgh, EH9 3FD, United Kingdom}

\author{Georg Rempfer}
\affiliation{Institute for Computational Physics, Universit{\"a}t Stuttgart, Allmandring 3, 70569 Stuttgart, Germany}

\author{Denis Botin}
\affiliation{Institut f{\"u}r Physik, Johannes Gutenberg Universit{\"a}t Mainz, 55128 Mainz, Germany}

\author{Christian Holm}
\affiliation{Institute for Computational Physics, Universit{\"a}t Stuttgart, Allmandring 3, 70569 Stuttgart, Germany}

\author{Thomas Palberg}
\affiliation{Institut f{\"u}r Physik, Johannes Gutenberg Universit{\"a}t Mainz, 55128 Mainz, Germany}

\author{Joost de Graaf}
\email{j.degraaf@ed.ac.uk}
\affiliation{SUPA, School of Physics and Astronomy, The University of Edinburgh, King's Buildings, Peter Guthrie Tait Road, Edinburgh, EH9 3FD, United Kingdom}

\title{Microfluidic Pumping by Micromolar Salt Concentrations}

\date{\today}

\begin{abstract}
An ion-exchange-resin-based microfluidic pump is introduced that utilizes trace amounts of ions to generate fluid flows. We show experimentally that our pump operates in almost deionized water for periods exceeding \SI{24}{\hour} and induces fluid flows of \si{\micro\meter\per\second} over hundreds of \si{\micro\meter}. This flow displays a far-field, power-law decay which is characteristic of two-dimensional (2D) flow when the system is strongly confined and of three-dimensional (3D) flow when it is not. Using theory and numerical calculations we demonstrate that our observations are consistent with electroosmotic pumping driven by \si{\micro\mole\per\liter} ion concentrations in the sample cell that serve as `fuel' to the pump. Our study thus reveals that trace amounts of charge carriers can produce surprisingly strong fluid flows; an insight that should benefit the design of a new class of microfluidic pumps that operate at very low fuel concentrations.
\end{abstract}

\maketitle

\section{\label{sec:intro}Introduction}

Fluid, solute, and colloid transport on the microscale pose a significant challenge, due to external pressure-driven pumping requiring the pump itself to withstand large forces. To circumvent this issue, a range of microfluidic pumps has recently been developed~\cite{andersson01, santiago04, kline05, paxton06, ibele07, chang07, nisar08, ibele09, jun10, shields10, hong10, hogg10, solovev11, zhang12, yadav12, reinmuller12, farniya13, sengupta14, esplandiu15, ortiz16, zhou16}, most of which exploit self-generated solute gradients. Typically, a solute-gradient-based (osmotic) pump consists of a source/sink of solute molecules, close to the surface of a sample cell. The solutes are produced/consumed either by chemical reactions on the surface of the pump~\cite{kline05, paxton06, ibele07, jun10, hong10, hogg10, solovev11, zhang12, yadav12, mcdermott12, farniya13, sengupta14, esplandiu15, ortiz16, zhou16, niu16b, velegol16}, the pump slowly dissolving~\cite{ibele09, mcdermott12, velegol16}, or exchange reactions taking place within the pump~\cite{reinmuller12, niu16b}. This sets up a concentration gradient in the fluid and along the surface of the sample cell. The interaction between the solutes and the surface causes a force on the fluid, which --- coupled with the spatial heterogeneity of the solutes --- leads to fluid flow, in a process that is referred to as osmosis~\cite{anderson89}.

In osmotic pumps, the solute thus acts as `fuel', which enables the pump to move fluid around. Such pumps generate relatively small forces applied over a much larger range of the fluid through long-ranged concentration gradients, thus overcoming the issues that face external pressure-driven pumps. Depending on the nature of the surface-solute interactions, neutral or Coulombic, the pump is either diffusioosmotic or electroosmotic. However, there is strong evidence that solute-density~\cite{ortiz16} and thermal-convection~\cite{sengupta14} effects can also play a role for large pumps. 

A fundamental problem for microfluidic pumping based on osmosis is the need for a solute (gradient) in the fluid medium that also contains the material to be transported, since solutes can interact with the transported material. For instance, pumps that utilize catalytic decomposition of hydrogen peroxide~\cite{kline05, paxton06, solovev11} or hydrazine~\cite{ibele07} are not biocompatible and these solutes will also react (unfavorably) with other materials. The working of pumps that instead employ enzymatic reactions to convert biomolecules, such as urease,~\cite{sengupta14,ortiz16} will be inhibited when material that is transported reacts with the solute itself. This also limits their use in transporting biological material, which typically interacts with such biomolecules. To solve this issue, it is desirable to design pumps that are driven by unreactive solutes, preferably in minimal amounts, and that have a limited impact on their environment.

In this paper, we introduce a microfluidic pump that accomplishes this goal. Our pump is experimentally shown to function in almost completely deionized water for periods of over \SI{24}{\hour}. We study the fluid flow by means of tracer velocimetry (close to the bottom of the sample cell) and show that the pumping speed is in the \si{\micro\meter\per\second} range over hundreds of \si{\micro\meter}. The dependence of this flow on the size of the pump and the added salt concentration in the system is also characterized. It is further experimentally demonstrated that solute-density and thermal-convection effects do not play a role in our system. We therefore hypothesize that our pump operates on trace amounts of ions present in the bulk fluid, by exchanging one species of ion for another, thereby generating a diffusion potential which drives electroosmotic flow. This sets it apart from other microfluidic pumps that generate flow by slow dissolution of the pump itself, see,~\textit{e.g.}, Refs.~\cite{ibele09, mcdermott12}. Specifically, our pump only modifies the identity of the ionic species in the bulk, whereas dissolving pumps increase the bulk ion concentration.

Furthermore, we show experimentally that the decay of the flow velocity can be modified by changing the geometry of the sample cell on the \si{\milli\meter} length scale. The far-field, power-law decay of the speed with the radial distance $r$ is either quasi-2D $(\propto r^{-1})$ for small cell heights ($\le \SI{2}{\milli\meter}$), or 3D $(\propto r^{-2})$ for tall cells ($\ge \SI{10}{\milli\meter}$). Even in the quasi-2D regime, our system displays almost time-independent (steady-state) fluid flow profiles. This is surprising, as 2D diffusive systems are not expected to exhibit steady-state solutions.

We interpret our experimental findings using a combination of the numerical finite-element method (FEM) and analytic calculations. It is shown that the experimental observations can indeed be understood by the resin exchanging trace amounts of cations from its surroundings with protons from its interior. We estimate the relevant trace cation concentration to be in the low micromolar range. The experimental observations are further shown to be consistent with an electroosmotic pumping mechanism: the difference in ion mobility between the protons and the exchanged cations sets up a diffusion potential that causes flow toward the exchange resin in the absence of a net electrical current. The mechanism is the same as previously found for similar ion-exchange pumps~\cite{reinmuller12} as well as dissolving pumps~\cite{ibele09, mcdermott12}. However, our results indicate that ion-exchange-resin-based microfluidic pumps have a surprisingly small lower bound to the ion concentration under which they can operate, which we chart in this paper. 

In our numerical work, we directly model the electroosmotic flow generated by ion exchange in the geometry of the experiment. We employ steady-state solutions for the concentration fields, electrostatic potential, and fluid velocity using the FEM. These computations go far beyond the thin electrostatic screening limit that is typically considered for such systems and give insight into the flow throughout the cell. Using analytic theory, we investigate the time dependence of the flow in the quasi-2D far field. We use our analytic theory to prove that the experimentally observed steady-state flow can be explained by the fact that the flow is driven by concentration gradients. While the relevant solute concentrations evolve over time and have no steady state, the concentration gradients become time-independent beyond a characteristic, system-dependent diffusion time that we identify. 

Finally, we can explain the scaling of our results with cell height in terms of interaction between the out-of-equilibrium ion fluxes and the confining geometry. Here, we observe qualitative, but not quantitative, agreement between the experiments and the numerical calculations. In the experiment, the power-law decay of the flow sets in unexpectedly close to the ion-exchange resin. We argue that this is due to the neglect of solute transport by advection in our calculations, which is necessary to make progress in both numerical and analytic theory. Accurately modeling the near-field effect of advection will be important to understanding the formation and performance of swimmers comprised of mobile ion-exchange resins and inert particles~\cite{niu16a} and therefore presents challenges for future study.

Our results on ion-exchange-resin-based microfluidic pumps lead to the startling finding that trace amounts of ions are sufficient to generate significant fluid flow, which is driven by diffusion-potential electroosmosis. This insight should prove instrumental for the design of new microfluidic pumps operating in close-to-deionized water, which is the natural and often desirable environment in which to perform experiments. It furthermore provides compelling evidence that the effect of small amounts of charge and minute ionic fluxes may have significant consequences in other systems, such as chemically self-propelled colloids.

\section{\label{sec:experiment}Experiments}

In this section, we describe the experimental setup for a single ion-exchange-resin pump and characterization of the tracer properties used in our velocimetry measurements. We also provide quantification of a wide range of resin pumps and tracers to show the generality of our findings. Finally, we study the impact of added salt on the pumping.

\subsection{\label{sub:tracer}Tracer Characterization}

Polystyrene (PS) tracers were used for the velocimetry (PIV) measurements of our ion-exchange-resin pump. Stock PS particle suspensions (Microparticle GmbH, Germany) were diluted with distilled water and thoroughly deionized using ion-exchange resin (Amberlite K306, Carl Roth GmbH~+~Co.~KG, Karlsruhe, Germany). The electrophoretic mobilities of the PS tracers were determined using micro-electrophoresis in a custom-built, disposable setup. For this setup, we employed a Perspex cell \SI{45}{\milli\meter} in height and with a square cross section (\SI{10}{\milli\meter\square} edge length). Based on the geometry of Uzgiris~\cite{uzgiris74, uzgiris81} two platinum electrodes of width \SI{1}{\milli\meter} were mounted into the center of the Teflon$^{\scriptsize{\mathrm{\textregistered}}}$ cap sealing the cell. This ensures sufficient electrode-wall distances to effectively reduce stray-field-driven electroosmosis at the cell walls. The electrode spacing was set to \SI{1}{\milli\meter} to obtain homogeneous electric fields and square-wave alternating voltages of $\pm \SI{1}{\volt}$ were applied by a function generator (PeakTeck 4060 by PeakTeck GmbH, Germany). The cell was mounted on the stage of a micro-electrophoresis instrument (Mark II, Rank Bros. Bottisham, Cambridge, UK) supplying ultramicroscopic illumination and particle tracks were imaged using exposure times of \SI{3}{\second} on a consumer digital single-lens reflex camera (DSLR; D800, Nikon, Japan).

\begin{table}[!htb]
\centering
\caption{\label{tab:mobility}The electrophoretic mobility $\mu_{\mathrm{E}}$ of the PS tracer particles used in this work. The first column gives the label, the second the diameter, and the third the value of $\mu_{\mathrm{E}}$.}
\begin{tabular}{|c|c|c|}
  \hline
  Type     & Diameter          & $\mu_{\mathrm{E}}$                                 \\
  -        & \si{\micro\meter} & $10^{-8}\:\si{\meter\squared\per\volt\per\second}$ \\
  \hline
  PS1      & $ 1.7\pm0.1$      & $-2.0\pm0.2$                                       \\
  PS7      & $ 7.6\pm1.0$      & $-2.6\pm0.3$                                       \\
  PS10     & $10.4\pm0.9$      & $-2.5\pm0.3$                                       \\
  PS15     & $15.2\pm0.9$      & $-2.5\pm0.2$                                       \\
  PS15COOH & $15.5\pm0.2$      & $-2.1\pm0.2$                                       \\
  \hline
\end{tabular}
\end{table}

Electrophoresis of the tracers was performed in the horizontal direction, while the particles sedimented in the vertical direction due to gravity. Thus the trace of a single particle has a saw-tooth shape. The mobility of the particles $\mu_{\mathrm{E}}$ was calculated from the averaged velocity in the horizontal direction $v_{\mathrm{E}}$ and is given by $\mu_{\mathrm{E}} = v_{\mathrm{E}}/E$, with $E$ the amplitude of the electric field. The obtained values of $\mu_{\mathrm{E}}$ are listed in Table~\ref{tab:mobility}. The values of $\mu_{\mathrm{E}}$ are relatively low due to the non-monotonic scaling of the electrophoretic mobility at low salt concentrations~\cite{midmore96}.

\subsection{\label{sub:resinpump}Velocimetry for the Ion-Exchange-Resin Pump}

\begin{figure}[!htb]
\centering
  \includegraphics[width=0.90\linewidth]{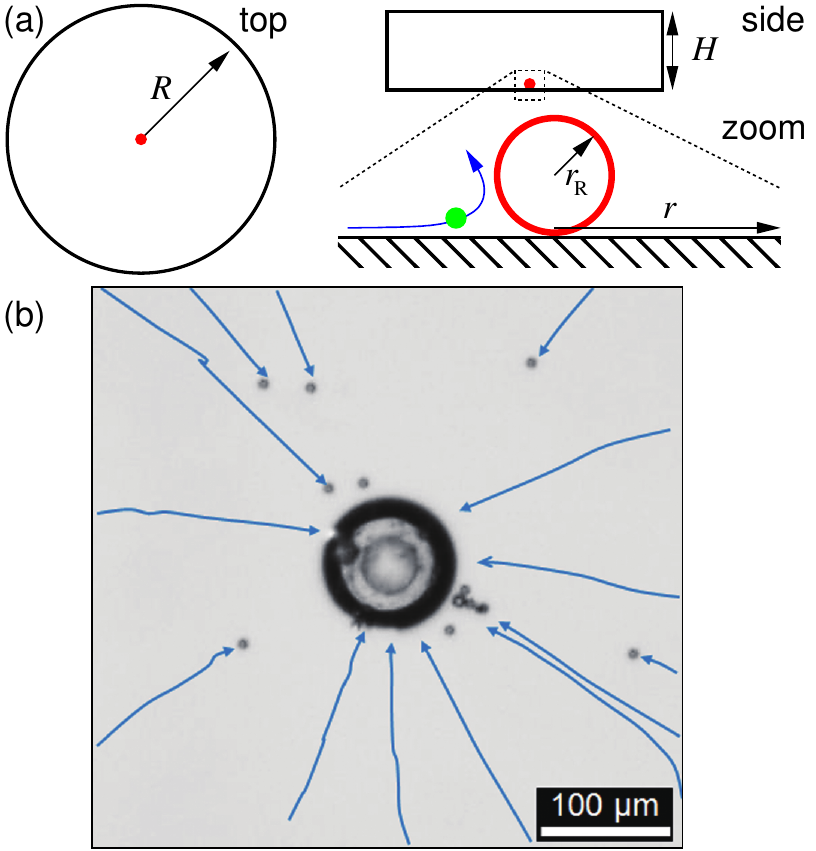}
  \caption{\label{fig:geom}The ion-exchange resin and sample cell. (a) Sketches of the geometry, showing top and side view of the cylindrical sample cell, with radius $R$ and height $H$. A zoom-in shows the exchange resin, with radius $r_{\mathrm{R}}$ in red, a polystyrene (PS) tracer in green flowing along a blue flow line. In this paper, radial distance $r$ is measured from the center of the resin. (b) Experimentally measured trajectories for the PS tracers (PS7) toward the resin (center of image), as shown in a top-view microscopy image. Blue arrows indicate the paths of the tracers, which are obtained from our image analysis.}
\end{figure}

For the characterization of the ion-exchange-resin pumps via tracer velocimetry, we constructed custom sample cells of radius $R = \SI{10}{\milli\meter}$ and several heights $H$ out of Poly(methyl methacrylate) (PMMA) rings attached to a microscopy glass slide and covered with another glass slide (soda lime glass of hydrolytic class 3 by VWR International), see the sketch in Fig.~\ref{fig:geom}a. The glass slides were washed with alkaline solution (Hellmanex$^{\scriptsize{\mathrm{\textregistered}}}$ III, Hellma Analytics) by sonication for \SI{30}{\minute}, then rinsed with tap water, and finally washed several times with doubly distilled water (distilled using a Quartz Hareaus Destamat; the conductivity was measured to be \SI{55}{\nano\siemens\per\centi\meter}). Spherical cationic resin beads (CGC50$\times$8, Purolite Ltd, UK; exchange capacity \SI{1.7}{\eq\per\liter}~\footnote{The \si{\eq} stands for amount of charge that the resin can exchange (here per liter of resin): $\SI{1}{\eq} = \SI{1}{\mole}$ of monovalent ions, $\SI{0.5}{\mole}$ of divalent ions, etc.}) with radii ranging from $10$ to $\SI{50}{\micro\meter}$ were carefully glued to the bottom glass slide with a tiny amount of two-component glue (UHU plus sofortfest, UHU GmbH, Germany), which was then set aside for \SI{24}{\hour} to allow the glue to completely solidify. One resin bead was glued in each sample cell.

\begin{figure}[!htb]
\centering
  \includegraphics[width=0.90\linewidth]{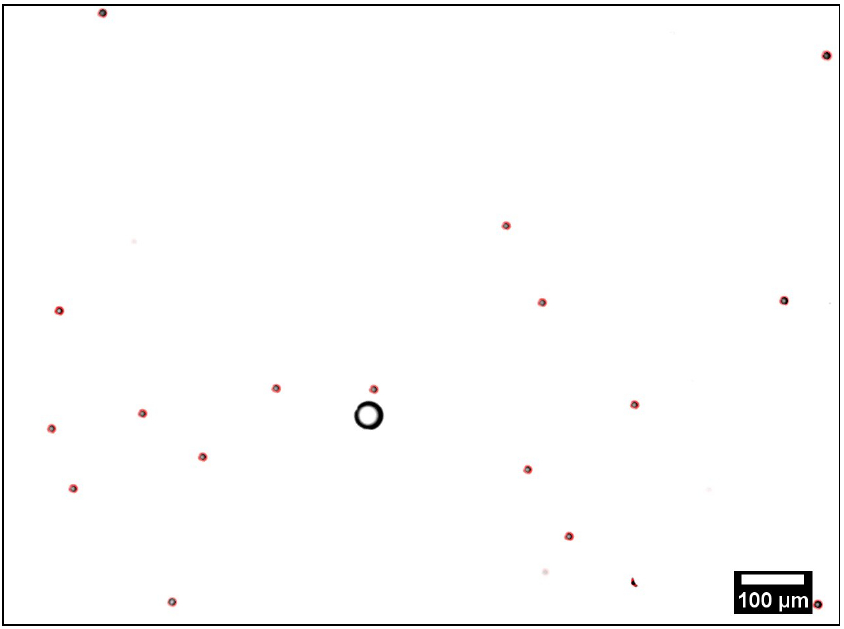}
  \caption{\label{fig:microscopy}Typical microscopy image of the tracers (red outlined circles) around the ion-exchange resin (black circle). The red outlines are obtained using the fit algorithm explained in the text.}
\end{figure}

The sample cell for the ion-exchange-resin pump experiments was loaded with a dilute PS-tracer suspension, prepared according to the above deionization procedure. It was subsequently mounted on the stage of an inverted scientific microscope (DMIRBE, Leica, Germany), and observed in bright field, typically at $5\times$ magnification. Images were shot with a DSLR and videos recorded with standard video equipment at frame sizes of $5.2$~Mpix and frame rates of $30$~fps. We imaged an area with cross-section of (typically) larger than $\SI{1000}{\micro\meter}$, slightly above the bottom glass plate, focusing on the average hovering height of gravitationally settled PS tracers, see Fig.~\ref{fig:microscopy}.

The resin bead glued to the glass slide displayed significant fluid pumping, as evidenced by the PS tracers moving toward the individual resin beads; Supplemental Movie ``Exchange\_Resin\_Pump.avi'' gives an example of this for tracers close to the resin. These tracers come in from far away along the substrate, move up vertically from the substrate close to the resin, then move radially away from the resin, subsequently sediment to the substrate away from the resin, and finally move back toward the resin along the substrate, leading to a recirculation of the tracer particles. Along their path the tracer speed varies as a function of $r$. The radial dependence of $U_{\mathrm{PS}}$ was determined from the tracer positions in successive frames of the recorded movies. These positions were extracted using an in-house \textsf{Python} code. In brief: the circular perimeter of each particle was extracted using standard edge-detection methods, and then fitted to a circle using the Hough transform~\cite{hough59}, implemented in the \textsf{OpenCV} function \textsf{HoughCircle}, see Fig.~\ref{fig:microscopy}. Tracer positions in consecutive frames were compared to determine radial velocity. The velocity of a given tracer species for a specific ion-exchange-resin bead size was measured for $80-100$ individual PS particles for each bead and the results averaged over some $40-50$ beads. 

We made use of the following expression to determine the tracer speed $U_{\mathrm{PS}}$
\begin{align}
\label{eq:travel} U_{\mathrm{PS}}(r) &= \left\langle \frac{\hat{r} \cdot \Delta \vec{s}}{\Delta t} \right\rangle ,
\end{align}
where $\hat{r}$ is the 2D unit vector pointing from the resin to the tracer, $\Delta \vec{s}$ is the displacement of the tracer between frames (time between frames $\Delta t$), ``$\cdot$'' is the inner product, and $\langle \cdots \rangle$ indicates averaging over all tracers that are a distance $r$ from the resin's center. 

\begin{figure}[!htb]
\centering
  \includegraphics[width=0.90\linewidth]{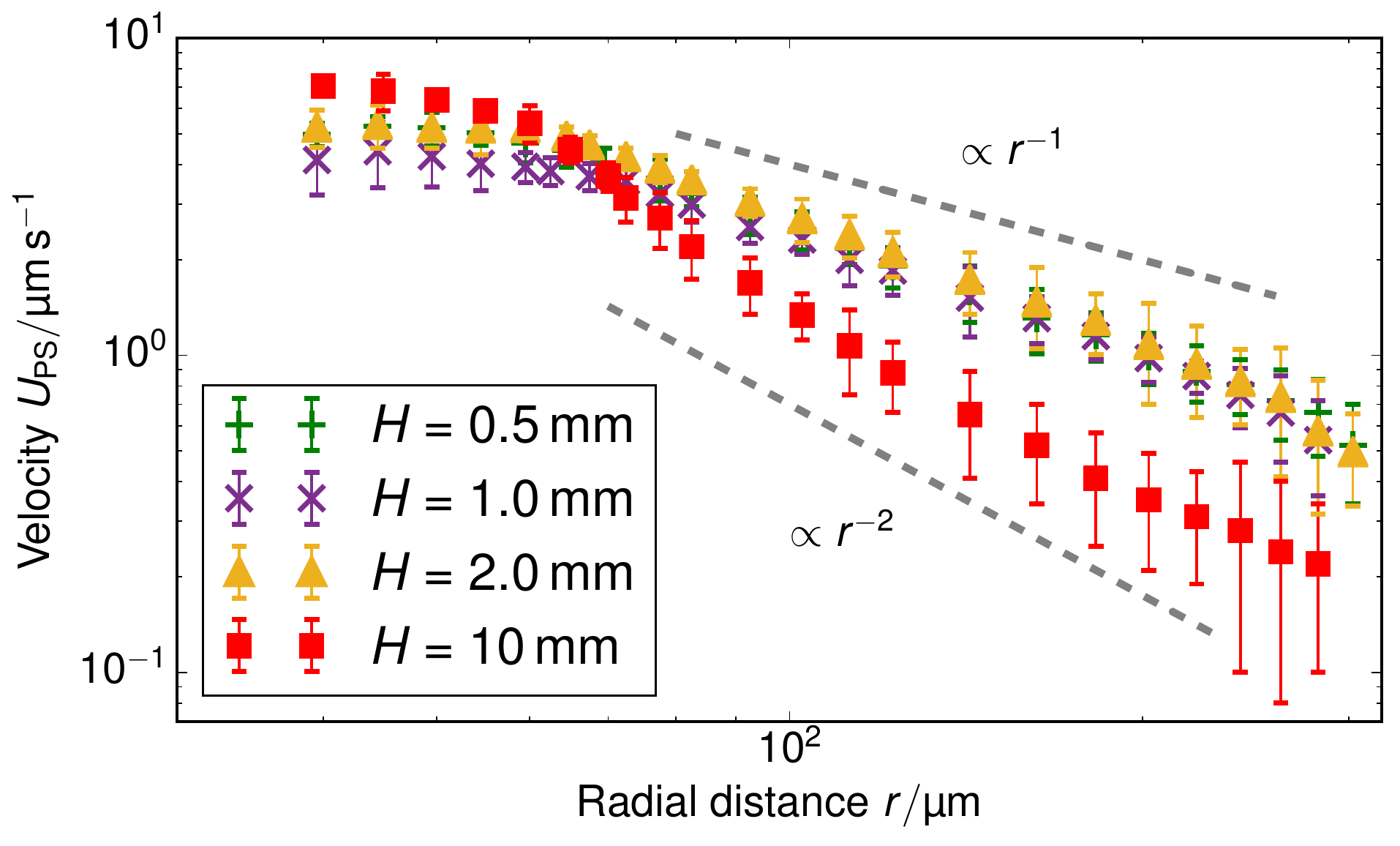}
  \caption{\label{fig:velo}The speed of the tracer $U_{\mathrm{PS}}$ as a function of the distance $r$ for several values of the sample cell height $H$, an ion-exchange resin with radius $r_{\mathrm{R}} = \SI{22.5}{\micro\meter}$, and PS7 tracers. The symbols show the experimentally measured values; the standard error is given for each data point. The gray dashed lines serve as a guides to the eye for the power-law decay.}
\end{figure}

The results of our velocimetry are shown in Fig.~\ref{fig:velo}, which provides $U_{\mathrm{PS}}$ as a function of the radial distance. Two regimes can be distinguished. For $r \lesssim \SI{75}{\micro\meter}$, there is a slight increase in the tracer speed, followed by a maximum and subsequent decrease (this is more evident in Fig.~\ref{fig:veltrac}). For $r \gtrsim \SI{75}{\micro\meter}$ the speed decreases with a power law and is appreciable over at least \SI{300}{\micro\meter}. For sample cells with a height of $H = \SI{1}{\milli\meter}$, we find that $U_{\mathrm{PS}} \propto r^{-0.9 \pm 0.1}$ in the far field ($H = \SI{0.5}{\milli\meter}$, $U_{\mathrm{PS}} \propto r^{-1.2 \pm 0.1}$; $H = \SI{2}{\milli\meter}$, $U_{\mathrm{PS}} \propto r^{-1.1 \pm 0.1}$), while for the sample with height $H = \SI{10}{\milli\meter}$, the fitted decay is $U_{\mathrm{PS}} \propto r^{-2.2 \pm 0.3}$. 

We concentrate on understanding the far-field regime throughout this paper, as in the near field there are several competing effects, including electrophoresis, local flow, and interaction with the substrate, which complicate understanding of the physics. For instance, it is difficult to assess on the strength of our experiments and the theory what the cause of the apparent near-field maximum in the tracer speed is. Fluid incompressibility could explain the decrease in speed close to the resin, \textit{i.e.}, an increasingly upward-directed component of the near-field flow requires a decrease in the horizontal component. However, other possibilities cannot be excluded at this time.

\subsection{\label{sub:general}Resin Size, Tracer Properties, and Salt Concentration}

\begin{figure}[!htb]
\centering
  \includegraphics[width=0.90\linewidth]{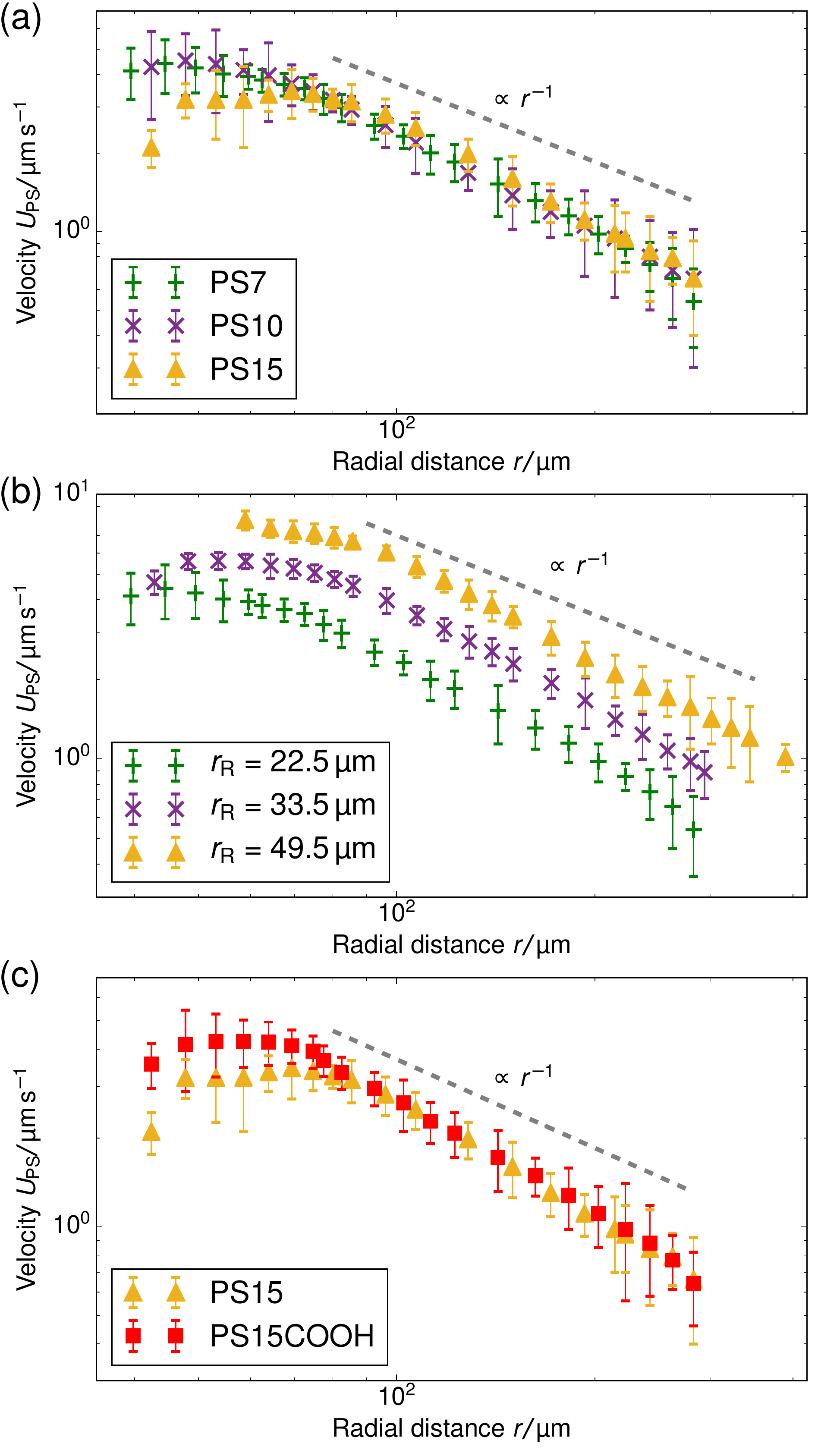}
  \caption{\label{fig:veltrac}Velocity of tracer particles $U_{\mathrm{PS}}$ as a function of radial distance $r$ for a cell height of $H = \SI{1.0}{\milli\meter}$. (a) Three different sized PS particles are pumped by resin beads with $r_{\mathrm{R}} = \SI{22.5}{\micro\meter}$. (b) Three different sized resin beads are used to form a pump, the tracer particle is PS7 for each. (c) Two PS tracers of the same size, but with different $\mu_{\mathrm{E}}$, are pumped by resin beads with $r_{\mathrm{R}} = \SI{22.5}{\micro\meter}$. In all panels, the gray dashed line serves as a guide to the eye for the power-law decay and the standard error is given for each data point.}
\end{figure}

In this section we demonstrate the generality of the fluid pumping by ion-exchange resins. The systematic quantification of the tracer speed $U_{\mathrm{PS}}$ as a function of $r$ is shown in Fig.~\ref{fig:veltrac}.

\begin{itemize}

\item In Fig.~\ref{fig:veltrac}a we vary the size of the PS tracers for tracers which have similar electrophoretic mobility. In the far-field region there is a power-law decay of the tracer speed, which is insensitive to the type of tracer used within the error bar. This shows that in the far-field the size of the tracer does not play a role.

\item Figure~\ref{fig:veltrac}b shows results for three resin sizes (radius $r_{\mathrm{R}}$). A larger resin induces a stronger electroosmotic flow over a larger range. We analyzed the far-field tracer speed by fitting the three curves using power-law decays. Then, we established the speed at an arbitrary far-field distance ($r = \SI{150}{\micro\meter}$) as a function of the size. For these three data points, we found a linear dependence through the origin $U_{PS}(r = \SI{150}{\micro\meter}) \approx r_{R} \times (5.5\pm0.5) \times 10^{-2}\:\si{\per\second}$; the offset $\approx \SI{0.15}{\micro\meter\per\second}$ is negligible --- similar scaling was observed for other far-field distances. This strongly indicates that the process is diffusion limited. By diffusion limited, we mean the upper speed limit imposed by the rate at which ions can diffuse towards the resin bead from the bulk reservoir. In this limit the flux through the particle surface $j_{s,\mathrm{dl}}$ (per unit area) is determined by the diffusivity of the ions $D^{\star}$ and the concentration far away $\rho^{\star}$, with the familiar diffusion-limit scaling $j_{s,\mathrm{dl}} \propto D^{\star} \rho^{\star} / r_{\mathrm{R}}$~\cite{brown16}. The speed is proportional to the total flux through the resin, \textit{i.e.}, $U_{\mathrm{PS}} \propto 4 \pi r_{\mathrm{R}}^{2} j_{s,\mathrm{dl}} \propto D^{\star} \rho^{\star} r_{\mathrm{R}}$, giving the linear dependence with $r_{\mathrm{R}}$ observed in the experiment.

\item In Fig.~\ref{fig:veltrac}c, we vary the electrophoretic mobility of the tracers, but not their size. It is evident that these tracer particles have the same velocity within the error bar in the power-law regime. This shows that the results are reproducible with nominally similar ($\mu_{\mathrm{E}}$ is comparable within the error bar), but possibly slightly different particles.

\end{itemize}

\begin{figure}[!htb]
\centering
  \includegraphics[width=0.90\linewidth]{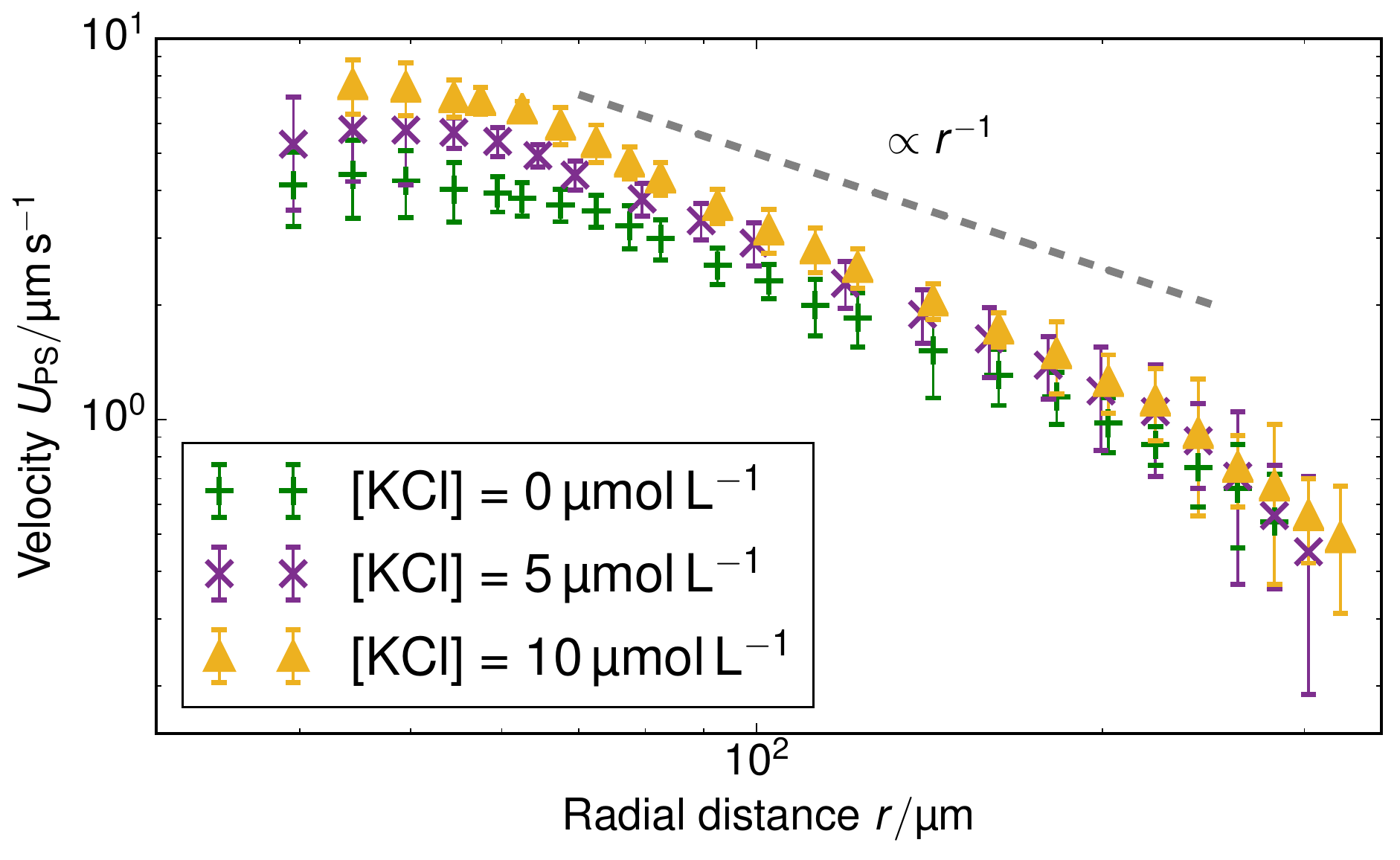}
  \caption{\label{fig:velsalt}Tracer speed $U_{\mathrm{PS}}$ as a function of radial distance $r$ for different added \ce{KCl} concentrations: no added salt (green plusses), \SI{5}{\micro\mole\per\liter} (purple crosses), and \SI{10}{\micro\mole\per\liter} (yellow triangles), for a cell with $H = \SI{1.0}{\milli\meter}$ and a resin with $r_{\mathrm{R}} = \SI{22.5}{\micro\meter}$. The gray dashed line serves as a guide to the eye for the power-law decay and the standard error is given for each data point.}
\end{figure}

Finally, we added \ce{KCl} solution (Merck KGaA, Germany) to the sample cell for $H = \SI{1}{\milli\meter}$ and the $r_{\mathrm{R}} = \SI{22.5}{\micro\meter}$ resin beads. Figure~\ref{fig:velsalt} shows the change in tracer speed: adding \SI{5}{\micro\mole\per\liter} \ce{KCl} increases $U_{\mathrm{PS}}$, adding \SI{10}{\micro\mole\per\liter} instead, increases the speed further. That is, a higher concentration of exchangeable ions induces stronger flow. However, at a \ce{KCl} concentration of \SI{80}{\micro\mole\per\liter}, the velocity of tracer particles is effectively zero (therefore not shown here). For the \SI{80}{\micro\mole\per\liter} sample, we also do not observe any Brownian motion of the tracer beads. This indicates that the beads have become firmly stuck to the sample cell wall, probably because of the increased electrostatic screening at this higher salt concentration. Therefore, we \textit{cannot} use this \SI{80}{\micro\mole\per\liter} data to infer a drop in pumping speed at higher salt concentration. This is in line with similar findings for chemically-propelled swimmers in Ref.~\cite{brown14}.

\subsection{\label{sub:invert}An Inverted Pump}

We inverted our setup to check whether solute density variations or thermal convection effects played a role in our system, as is the case in Refs.~\cite{ortiz16, sengupta14}. That is, we glued the resin to the top glass slide and examined the movement of the tracers. In order to ensure that the tracers were at the top cover slide, we modified the overall density of the solution by adding glycerol (water:glycerol mass ratio of 1:0.3) to slightly exceed the density of our PS particles. We used tracers with a diameter of \SI{3.3}{\micro\meter} here and we increased the size of the ion-exchange resin to $r_{\mathrm{R}} = \SI{250}{\micro\meter}$, in order to increase the speed of the tracers in this mixture of higher viscosity (approximately double that of water). 

Supplemental movie ``Inverted\_Resin\_Pump.avi'' shows the result of this experiment. It is clear that inverting the pump did not change the direction in which the tracers move toward the resin. While we increased the overall density of the mixture, this should not affect the possible density variations induced by ion exchange. Our experiment therefore rules out density variation effects. 

\section{\label{sec:consider}General Considerations}

We performed a theoretical/numerical analysis of the pump to gain understanding of the fluid flow observed in our experiments and to show that the observed fluid pumping is indeed caused by trace amounts of cations in the sample cell. In this section, we start with several general considerations to provide a background for our calculations.

\begin{itemize}

\item The resin is designed to exchange \ce{H+} for other cations in the bulk, with a capacity of \SI{1.7}{\eq\per\liter}. There are \textit{no} decomposition-type surface reactions. Nor does our pump itself dissolve, as is the case in Refs.~\cite{ibele09, mcdermott12}.

\item The surface of the sample cell is charged and we measured the zeta potential of the bottom glass slide to be $\zeta \approx -(105 \pm 5)\:\si{\milli\volt}$~\footnote{We used Doppler velocimetry with PS tracers~\cite{palberg12}, originally designed to measure bulk electrokinetics in colloidal suspensions, to determine the electroosmotic mobility $\mu_{\mathrm{wall}}$ of cleaned glass slides using a custom made cell with exchangeable sides for the top and bottom. Standard electrokinetic theory was used to calculate the zeta potential from the mobility~\cite{delgado07}.}. The negative surface charge is due to the dissociation of surface groups, which release cations into the bulk. However, because the sample cell is filled with deionized water and carefully rinsed before preparation, there will be very few non-protonic cations present. 

\item Dissolved \ce{CO2} forms carbonic acid and thus creates cations in the form of protons that screen the wall charge. We measured a $\mathrm{pH} \sim 5.4$ at the onset of the experiment, consistent with typical $\mathrm{pH}$ values for water in equilibrium with atmospheric \ce{CO2}~\cite{millero95}. The cations associated with \ce{CO2} dissociation (protons) are the same as the ions inside the ion-exchange resin, so they cannot contribute to electroosmosis via ion exchange.

\end{itemize}

One might assume that the exchange resin cannot exchange protons for other cations, as any non-protonic cations released from the cell walls will have been washed away during preparation of the sample cell, according to our second point. This would then prevent ion-exchange-based electroosmosis. 

However, we will argue that a very low concentration of cations remaining in the bulk after cleaning would be sufficient to fuel pumping. We estimate here the residual concentration of ions that would be required in this case. During a period of \SI{24}{\hour}~\footnote{We experimentally measured a tracer speed decrease of only a factor $2$ over a \SI{24}{\hour} period, justifying the assumption of almost constant pumping. The shape of the speed profile remained unchanged.}, a spherical pump of $r_{\mathrm{R}} = \SI{22.5}{\micro\meter}$ in radius exchanges at most a part of its ionic content \textit{via} an (assumed) constant surface flux density $j_{s}$. Let us further assume 25\% of the original content to account for an unmodified pumping speed over the course of the experiment. Then it follows from the resin size and exchange capacity that $j_{s} \lesssim \SI{5e-8}{\mole\square\per\meter\per\second}$. This value is reasonable, as similar numbers are found for self-electrophoretic Janus swimmers that move at speeds comparable to our maximum $U_{\mathrm{PS}}$~\cite{brown16}. To make this level of exchange possible, the non-protonic cation concentration in the sample cell has to be at least \SI{1e-7}{\mole\per\liter} --- the total number of exchanged ions over \SI{24}{\hour} and the volume of the sample cell were used to arrive at this number. Again, we can assume only a fraction of the total ions present are exchanged. This would lead to an estimate for the cation concentration of $\rho^{\star} = \SI{1}{\micro\mole\per\liter}$ (if 10\% is exchanged). Such low ion concentrations could be contributed by ``impurity'' cations released from the glass slides into the bulk fluid following rinsing. Taking $\rho^{\star}$ and a typical value for cation diffusivity of $D^{\star} \approx \SI{2e-9}{\meter\squared\per\second}$, the surface flux in the diffusion-limited regime is $j_{s,\mathrm{dl}} \approx D^{\star} \rho^{\star} / r_{\mathrm{R}} = \SI{4e-8}{\mole\square\per\meter\per\second}$ (using $\ce{K+}$ for the cationic contaminant, which has a typical cation diffusivity). Our estimate for $j_{s}$ is thus in the physically reasonable regime, close to the diffusion limit, in accordance with our experimental result.

This proposed mechanism of generating fluid flow by ion exchange is as follows. Exchanged protons moving away from the resin have a higher mobility than the to-be-exchanged cations moving toward the resin --- \ce{H+} has the highest mobility of any ion --- and thus the protons have a greater \textit{diffusive} flux. To prevent bulk charge separation due to the difference in diffusivity, an electric field ($E$-field) is formed to compensate for the difference in diffusive flux with a migrative flux (\textit{via} the $E$-field). The $E$-field points toward the resin and prevents charge separation in the bulk, by slowing down the \ce{H+} and accelerating the cations, such that the total fluxes remain equal and opposite throughout. Since the associated electrostatic potential stems from a difference in ion diffusivity (equivalently mobility) this mechanism is ``diffusion potential'' based~\cite{anderson89,reinmuller12}. The $E$-field acts on all ions in the system. However, the $E$-field only exerts a significant force on the fluid in a screening layer close to the chamber boundaries, where there is an excess of cations. In the electrically neutral bulk, the $E$-field has a vanishingly small effect. The result is that the $E$-field drives fluid flow along the glass slides towards the resin, which then, through incompressibility, generates a backflow outwards along the horizontal center-plane of the chamber.

We verified that this simple picture and our estimates can indeed give rise to the observed flow speed and direction using finite-element-method (FEM) simulations representative of the experimental geometry, see Sections~\ref{sec:simulation} and~\ref{sec:FEMres}. We also used linearized, analytic theory to study the regime where the geometry can be considered as quasi-2D,~\textit{i.e.}, for $r > H$, see Section~\ref{sec:theory}. In both cases, we solved the associated time-dependent electrokinetic equations, which we discuss next.

\section{\label{sec:electrokinetic}Electrokinetic Equations}

To model the electroosmotic flow around the ion-exchange resin, we require three coupled equations, collectively known as the electrokinetic equations: Nernst-Planck for the solutes, Poisson for the electrostatics, and Stokes for the fluid flow, together with boundary conditions for the respective problems. We explain the three equations in detail below and discuss the boundary conditions for the simulations in Section~\ref{sec:simulation} and the approximations made for the theory in Section~\ref{sec:theory}, respectively.

The \textbf{Nernst-Planck equation} describes the diffusion and migration of the solute species. Here, we consider three ionic solute species in the fluid, protons $\ce{H+}$, potassium $\ce{K+}$, and chloride $\ce{Cl-}$. The protons are loaded in the exchange resin and the choice of the two other ions is arbitrary. Here, we selected two ions with almost equal diffusivities $D_{\ce{K+}} \approx D_{\ce{Cl-}}$~\cite{harned49} to avoid the complication of additional (but relatively small) diffusion potentials. These three species are indexed by $i \in \{\ce{H+},\ce{K+},\ce{Cl-}\}$, respectively. We write $\rho_{i}$ for the time- and space-dependent concentration fields and $D_{i}$ for the molecular diffusivities. Then the flux of each species is given by
\begin{align}
	\label{eq:flux} \vec{j}_{i} &= - D_{i} \vec{\nabla} \rho_{i} + \vec{u} \rho_{i} - \frac{e z_{i}}{k_{\mathrm{B}} T} D_{i}  \rho_{i} \vec{\nabla}  \Phi, 
\end{align}
where $\vec{u}$ is the fluid velocity (accounting for advection), $k_{\mathrm{B}}$ is Boltzmann's constant, $T$ the temperature, $e$ the elementary charge, $z_{i}$ the valency, $\Phi$ the electrostatic potential, and $\vec{\nabla}$ the gradient operator. The continuity equation is given by
\begin{align}
	\label{eq:NP} \partial_{t} \rho_{i} &= - \vec{\nabla} \cdot \vec{j}_{i}, 
\end{align}
where $\partial_{t}$ denotes the time derivative. For the steady-state problem $\partial_{t} \rho_{i} = 0$. 

Before we move on to the other equations, we should comment on two simplifying assumptions typically made in the above description. 

\begin{itemize}

\item We ignored the advective contribution to the flow in Eq.~\eqref{eq:flux} in all our calculations. We consider the P{\'e}clet number, which give the ratio of diffusion to advection, to examine whether this is reasonable. A simple estimate is as follows: using a typical length scale of $H$ for the development of fluid flow in this problem, a typical non-protonic ion diffusivity of $D \approx \SI{2.0e-09}{\meter\squared\per\second}$~\cite{harned49}, and a typical speed of $\bar{U}_{\mathrm{PS}} \approx \SI{1.0}{\micro\meter\per\second}$, we arrive at $\mathrm{Pe} = \bar{U}_{\mathrm{PS}} H / D \approx 5$. This indicates that the value of $\mathrm{Pe}$ is probably high~\footnote{We believe that in practice the $\mathrm{Pe}$ number is likely to be self-limiting to a value $\approx 1$, as found for chemically propelled swimmers~\cite{cordova08}. This is because the high-concentration-gradient region centered around the colloid would be expelled into the bulk of the channel by a strong advective current --- see the direction of the flow lines in Fig.~\ref{fig:trajectories} --- where it would no longer contribute strongly to electrophoretic flow generation.}, so the advective term should not be ignored in Eq.~\eqref{eq:flux}. However, due to the computational complexity of our FEM calculations~\footnote{Making the $\mathrm{Pe} = 0$ assumption allows us to split the solute and solvent problems and solve them in series, rather than in parallel, leading to a strong reduction in the required mesh resolution and therefore of computer memory.}, as well as the need to linearize our analytic theory, this approximation \textit{must} be made in order to make progress. As we will see, the understanding of the physics of the resin pump is not strongly affected by this reduction.

\item We have ignored bulk ionic association-dissociation reactions, as described in Ref.~\cite{brown16}, which would have entered on the right-hand side of Eq.~\eqref{eq:NP} as coupled chemical source and sink terms. In the physical system, bulk exchange will lead to coupling of the $\ce{H+}$ flux coming from the ion-exchange resin and $\ce{H2O}$ and $\ce{OH-}$ present in solution via $\ce{H2O <=> H+ + OH-}$~\footnote{There will be similar association-dissociation reactions involving dissolved $\ce{CO2}$ in water, as well as other species. The arguments provided here apply equally to these reactions.}. The main effect of these bulk reactions would be to replace the relevant diffusion rate $D_{\ce{H+}}$ with an effective rate
\begin{align}
  \label{eq:Dstar} D_{\mathrm{av}} &\equiv \frac{D_{\ce{H+}} \rho_{\ce{H+}}^{\infty} + D_{\ce{OH-}} \rho_{\ce{OH-}}^{\infty}}{\rho_{\ce{H+}}^{\infty} + \rho_{\ce{OH-}}^{\infty}},
\end{align}
with $\rho_{i}^{\infty}$ the concentration ``very far'' away from the resin~\cite{brown16}. For the experimental $\mathrm{pH} \lesssim 5.4$, $[\ce{OH-}] \ll [\ce{H+}]$, so $D_{\mathrm{av}} \approx D_{\ce{H+}}$, to within 2\%. Hence, it is justified to ignore the bulk reactions here.

\end{itemize}

The electrostatic potential fulfills the \textbf{Poisson equation}, which is given by
\begin{align}
	\label{eq:Poisson} -\epsilon_{0} \epsilon_{\mathrm{r}} \nabla^{2} \Phi &= e \sum_{i} z_{i} \rho_{i}, 
\end{align}
where $\epsilon_{0}$ is the vacuum permittivity and $\epsilon_{\mathrm{r}}$ the (spatially constant) relative permittivity. It should be noted that the $\rho_{i}$ and $\Phi$ in the Poisson equation are time dependent and these two quantities provide the coupling between the Nernst-Planck and Poisson equations. For completeness, we introduce the electrostatic screening (Debye) length $\kappa^{-1}$ here via
\begin{align}
 \label{eq:Debye} \kappa^2= \frac{e^{2}}{\epsilon_{0} \epsilon_{\mathrm{r}} k_{\mathrm{B}} T} \sum_{i} z_{i}^{2} \rho_{i}^{\infty} .
\end{align}

Finally, we have the incompressible \textbf{Stokes equations} to describe the fluid flow. These read
\begin{align}
	\label{eq:stokes} \eta \nabla^{2} \vec{u} &= \vec{\nabla} p - \vec{f}; \\
	\label{eq:incompressibility} \vec{\nabla} \cdot \vec{u} &= 0, 
\end{align}
with $\eta$ the viscosity of the fluid, $p$ the hydrostatic pressure, and $\vec{f}$ the body-force density. Here, $\vec{u}$ and $\vec{f}$ are time-dependent quantities. Given the electrostatic potential and the densities of all (charged) species, we can specify the body-force density on the fluid to close the problem as 
\begin{align}
	\label{eq:closure} \vec{f} &= \sum_{i} \frac{k_{\mathrm{B}} T}{D_{i}} \vec{j}_{i} . 
\end{align}
This expression was obtained by first-order expansion of the chemical potential around thermodynamic equilibrium, which gives the gradient of the chemical potential as a driving force~\cite{rempfer16a}. The specific choice of this driving force is to eliminate the spurious flow due to inexact cancellation of pressure and electrostatic interactions in FEM calculations. It is, however, completely equivalent~\cite{rempfer16a} to the more commonly used expression
\begin{align}
  \label{eq:thclo} \vec{f} &=  e \sum_{i} z_{i} \rho_{i} \vec{\nabla} \Phi.
\end{align}
The only difference between Eqs.~\eqref{eq:closure} and~\eqref{eq:thclo} is the interpretation of the hydrostatic pressure: Eq.~\eqref{eq:closure} does not, while Eq.~\eqref{eq:thclo} does include the ideal-gas contribution from the dissolved solutes~\cite{rempfer16a}.

\section{\label{sec:simulation}Finite-Element Model of the Pump}

In this section we describe the boundary conditions for the above equation system and the specific choices made for the FEM modeling. Throughout, we used \textsf{COMSOL Multiphysics}$^{\scriptsize{\mathrm{\textregistered}}}$ \textsf{Solver 5.2a} to numerically solve the electrokinetic equations for a model setup of the experimental geometry. 

\begin{figure}[!htb]
\centering
  \includegraphics[width=0.90\linewidth]{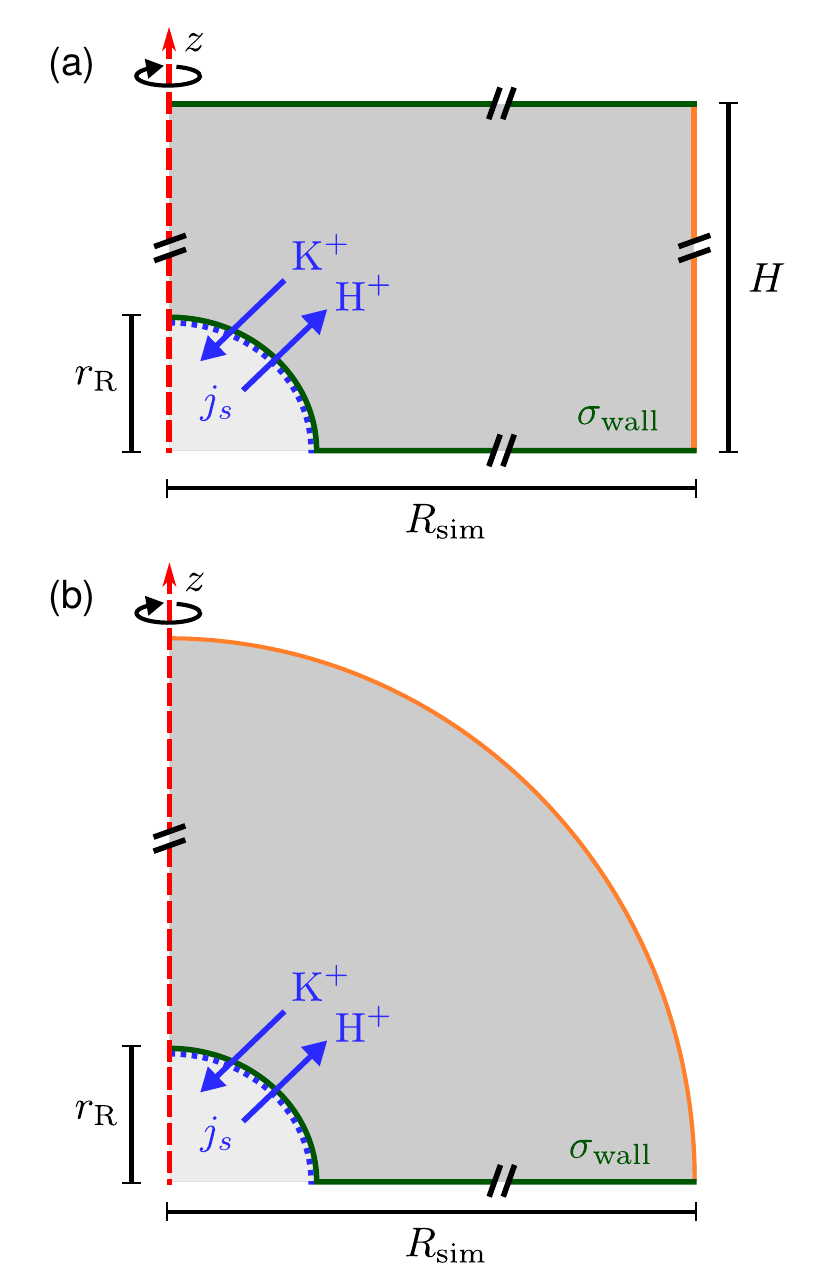}
  \caption{\label{fig:fem_setup}Schematic of the rotationally symmetric (a) top-bound and (b) ``unbound'' simulation domain, the red dashed line shows the symmetry axis. The resin is modeled as a hemisphere of radius $r_{\mathrm{R}}$ (lower-left corner). Cation exchange is modeled by an inward/outward directed flux $j_{s}$ of cations ({\ce{K+}}) and protons ({\ce{H+}}), respectively, see Eq.~\eqref{eq:cases}. A constant surface charge density $\sigma_{\mathrm{wall}}$ is imposed on the bottom (and top wall) and on the resin. All solid surfaces form no-slip boundaries for the hydrodynamics. The right-most boundary (orange line (a) or a circular arc (b)) is an ``open boundary'' for the hydrodynamic problem and we impose a pre-computed electrostatic profile and ion distributions on it, as explained in the text. Cut lines are used to emphasize that the domain is much larger than the resin, see Fig.~\ref{fig:mesh}.}
\end{figure}

We considered a 3D cylindrical portion of the microscopy cell, with the resin located on the symmetry axis of the cylinder. Due to the rotational symmetry of our setup, which corresponds closely to the experiment, the simulations could be performed on a quasi-2D axisymmetric domain, see Fig.~\ref{fig:fem_setup}. We considered two domains in order to simulate small sample heights $H \lesssim \SI{1}{\milli\meter}$ (a) and large sample heights $H \gtrsim \SI{2}{\milli\meter}$ (b) for the steady-state electrokinetic equations; we will come back to this in Section~\ref{sec:FEMres}. The latter domain is a half-open domain, which we will refer to as ``unbound''.

\begin{figure}[!htb]
\centering
  \includegraphics[width=0.90\linewidth]{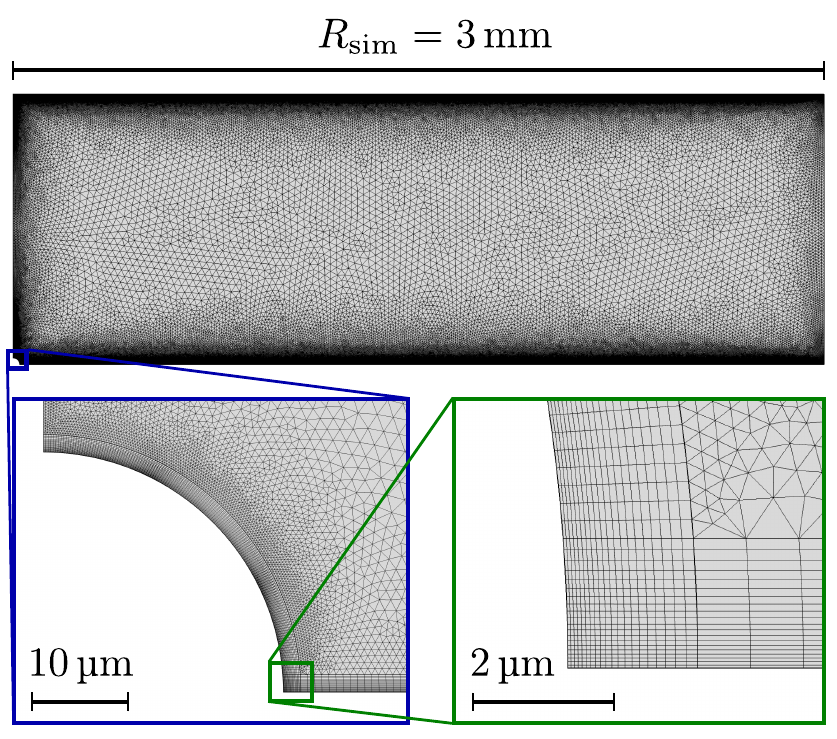}
  \caption{\label{fig:mesh}Example of the fine mesh used for our FEM calculations ($H=\SI{1}{\milli\meter}$). Close to the electrically charged surfaces (within 6 Debye lengths) quadrilateral elements are used, see insets. The rest of the domain is composed of triangular mesh elements.}
\end{figure}

Let us first describe the simulation domain that most accurately represents the experiment, see Fig.~\ref{fig:fem_setup}a. The bottom and top of the simulation domain correspond to the glass slides of the sample cell, the height of the sample cell $H$ is fully resolved. The radius of the simulated geometry is $R_\mathrm{sim}$. The spherical resin (experiment) is modeled as a hemisphere of radius $r_{\mathrm{R}}$ attached to the lower boundary (substrate). We chose a hemi-spherical resin, rather than a fully spherical one --- as in the experiment --- for simulation convenience. Specifically, the choice of a hemispherical resin facilitates the use of quadrilateral elements for the mesh, see inset to Fig.~\ref{fig:mesh}. This meshing would not be possible for a resin sphere in contact with the substrate, as is likely the case in the experiment, due to the cusp-like feature present in that geometry.

Quadrilateral elements are necessary, since we use the spurious-flow reducing method of Refs.~\cite{rempfer16b, rempfer16c, kreissl16}. This method consists of finely meshing several Debye lengths (in our case 6) using such elements around the walls and the rest of the domain using triangular elements, see Fig.~\ref{fig:mesh}. Quadrilaterals have the advantage that larger aspect ratios are permitted than for triangles, before numerical instabilities become important in FEM. It should be noted that our choice of a hemispherical resin will only affect the near-field flow around the resin. Beyond a certain distance, the resin can be considered a point source for protons and sink for potassium ions, respectively, and the details of its shape thus become irrelevant. This far field is the regime of interest to us.

At the edge of the domain, there is an ``open boundary'' for the hydrodynamic problem. This implies that there is no fluid momentum flux through the boundary. Since there is no convective momentum transport in the Stokes equations, there can be flow, but no stress normal to the boundary. This is a standard technique to model a piece of a domain that is embedded in a larger physical region, without modeling the full geometry, but allowing for the flow lines not to be closed within the domain. The unbound simulation domain (Fig.~\ref{fig:fem_setup}b) is the same as the top-bound domain, but replaces the top glass slide with a hemispherical (open-boundary) domain. 

We now provide the expressions for the boundary conditions used in the FEM model.

\begin{itemize}

\item For all solute species, no-penetration conditions are imposed in the \textbf{Nernst-Planck equation} on the bottom/top of the cell
\begin{align}
	\label{eq:noflux} j_{s} \equiv \left. \hat{n} \cdot \vec{j}_{i} \right\vert_{r_{\mathrm{R}}} &= 0 , 
\end{align}
where $\hat{n}$ is the unit normal to the boundary pointing into the fluid. 

For the resin, we only impose no-penetration conditions for $\ce{Cl-}$. The exchange of $\ce{H+}$ and $\ce{K+}$ is modeled via out- and influx on the resin, respectively. To be precise, we impose the following flux boundary condition
\begin{align}
	\label{eq:cases} j_{s} \equiv \left. \hat{n} \cdot \vec{j}_{i} \right\vert_{r_{\mathrm{R}}} &= 
    \begin{cases}
    + k_{\mathrm{ex}} \rho_{\ce{K+}} & i=\ce{H+} \\
    - k_{\mathrm{ex}} \rho_{\ce{K+}} & i=\ce{K+} \\
                                  0  & i=\ce{Cl-} 
    \end{cases} , 
\end{align}
where $k_{\mathrm{ex}}$ is the ion-exchange rate coefficient, which we need to determine by fitting to the experimental data. Note that we have assumed that the exchange is determined entirely by the cation concentration close to the resin. This is probably valid as long as the \ce{H+} concentration inside the resin is much larger than the cation concentration outside, which is the case for a fresh resin.

At the outer edge of the domain, the orange line in Fig.~\ref{fig:fem_setup}a, we impose concentration profiles for the ions that are based on the Poisson-Boltzmann solution for a two plate geometry with height $H$ and surface charge $\sigma_{\mathrm{wall}}$. For the geometry of Fig.~\ref{fig:fem_setup}b, the solution to the Poisson-Boltzmann equation for a single plate was used.

\item For the \textbf{Poisson equation}, we impose a constant surface charge density $\sigma_{\mathrm{wall}}$ on all solid surfaces via
\begin{align}
  \label{eq:surf} \hat{n} \cdot \vec{\nabla} \Phi &= - \frac{\sigma_{\mathrm{wall}}}{ \epsilon_{0} \epsilon_{\mathrm{r}} },
\end{align}
where the electrostatic potential is assumed to be evaluated at the boundary. The surface charge density is obtained from the experimental zeta-potential measurement through the Grahame equation~\cite{grahame47}
\begin{align}
  \label{eq:Grahame} \sigma_{\mathrm{wall}} &= \sqrt{ 4 \epsilon_{0} \epsilon_{\mathrm{r}} k_{\mathrm{B}} T \sum_{i=0}^{2} \rho_{i}^{\infty} } \textrm{sinh} \left( \frac{e \zeta }{k_{\mathrm{B}} T} \right) ,
\end{align}
with $\zeta$ the zeta potential. Note that it is not clear what the most appropriate boundary conditions are for the resin surface, hence we chose the same boundary condition (Eq.~\eqref{eq:surf}) as on the other surfaces for computational convenience.

\item The boundary conditions for the incompressible \textbf{Stokes equation} are no-slip boundary conditions on all solid surfaces ($\vec{u} = 0$), that is the bottom and top of the cell, plus the resin. At the outer edge of the domain, a no-normal stress boundary condition is applied, which reads
\begin{align}
	\label{eq:nostr} \left[ \eta \left( \vec{\nabla} \vec{u} + \left( \vec{\nabla} \vec{u} \right)^\mathrm{T} \right) - p \mathbb{I} \right] \cdot \hat{n} = 0, 
\end{align}
with ${}^\mathrm{T}$ denoting transposition and $\mathbb{I}$ the identity matrix.

\end{itemize}

A final detail for the FEM solver is that we used polynomial ansatz functions of order $2$ for the electrostatic, order $2$ for the diffusion/migration, and order $3 + 2$ for hydrodynamic equations. This is necessary in order to further reduce spurious flows. Despite these measures, as well as decoupling the solute and solvent problems by our low $\mathrm{Pe}$ assumption, extremely fine meshes are required, see Fig.~\ref{fig:mesh}, that push the boundaries of modern computational platforms in order to obtain convergence and sufficiently smooth results within a reasonable time.

\subsection{\label{sub:traject}Tracer Speed}

We determined the tracer speed from the solution of the above system of electrokinetic equations with boundary conditions as follows. The speed
\begin{align}
\label{eq:EOM} U_{\mathrm{PS}} &= \vec{u}\cdot\hat{r} + \mu_{\mathrm{E}} \vec{E}\cdot\hat{r},
\end{align}
is comprised of an advective term, which is captured by $\vec{u}$, and a component deriving from the electrophoretic mobility $\mu_{\mathrm{E}} \vec{E}$. We evaluate the velocity and electric field at a constant ``equilibrium height'', $h^{\star}$, where gravity balances electrostatic repulsion from the wall. Throughout, we used a constant height of $h^{\star} = \SI{5}{\micro\meter}$. The exact height in the experiment is difficult to measure, presumably varies locally, and changes with the environment. We therefore varied $h^{\star}$ between \SI{4}{\micro\meter} and \SI{10}{\micro\meter} to check how our specific choice affected the result. The resulting speed profiles turned out to be virtually the same in this range. This is because the fluid flow velocity, which is the major component in the tracer speed, varies over a typical length scale of order $H \gg h^{\star}$. Note that Eq.~\eqref{eq:EOM} treats the tracer particle as if it were a point-like object,~\textit{i.e.}, it does not perturb the flow and electric fields by its presence. In general we found that including the second term in Eq.~\eqref{eq:EOM} does not significantly modify the $U_{\mathrm{PS}}$, leading us to conclude that advection indeed dominates over electrophoretic effects for the tracer motion.

\subsection{\label{sub:fem_params}Parameter Choices}

We made the following parameter choices to simulate the experimental system. For the geometry of the simulation setup we typically chose: $r_{\mathrm{R}} = \SI{25}{\micro\meter}$ for the radius of the resin and $R_\mathrm{sim} = \SI{3}{\milli\meter}$ for the radius of the cylindrically symmetric domain. This choice was a compromise between the size of the sample cell, which is too large to numerically simulate in its entirety, and a domain size on which the power-law decay in the fluid velocity was observable in the steady-state FEM calculations. The height of the domain was chosen to match the relevant experimental setup, \textit{e.g.}, $H = \SI{1}{\milli\meter}$, with the open simulation domain representing the $H = \SI{10}{\milli\meter}$ domain, as we explain in Section~\ref{sec:FEMres}.

The fluid represents water at room temperature ($T = \SI{298.15}{\kelvin}$), which has a mass density of $\rho_\text{f} = \SI{1.0E3}{\kilogram\per\meter\cubed}$, viscosity $\eta = \SI{8.9E-4}{\pascal\second}$, and relative permittivity $\epsilon_{\mathrm{r}} = 78.4$. The diffusivities of the ionic species are $D_{\ce{H+}} = \SI{9.3e-09}{\meter\squared\per\second}$~\cite{haynes13} and $D_{\ce{K+}} = D_{\ce{Cl-}} = \SI{2.0e-09}{\meter\squared\per\second}$~\cite{harned49}. The bulk concentration of impurities was chosen to be $\rho_{\ce{K+}}^{\infty} = \SI{1.0}{\micro\mole\per\liter}$, in line with our estimates from the experiment, and $\rho_{\ce{H+}}^{\infty} = \SI{0.1}{\micro\mole\per\liter}$ ($\mathrm{pH} = 7$) or $\rho_{\ce{H+}}^{\infty} = \SI{4.0}{\micro\mole\per\liter}$ ($\mathrm{pH} = 5.4$), with $\rho_{\ce{Cl-}}^{\infty} = \rho_{\ce{K+}}^{\infty} + \rho_{\ce{H+}}^{\infty}$. The ion-exchange rate coefficient $k_{\mathrm{ex}} = \SI{3.08E-6}{\meter\per\second}$ was obtained by fitting the near-field velocity to the experiment for $H = \SI{1}{\milli\meter}$ in Fig.~\ref{fig:velo}. The surface charge density $\sigma_{\mathrm{wall}} = \SI{-4.03E-4}{\coulomb\per\meter\squared}$ was computed from the experimentally measured zeta potential $\zeta \approx \SI{-0.1}{\volt}$ using the Grahame equation~\eqref{eq:Grahame}~\cite{grahame47}. 

\section{\label{sec:FEMres}Finite-Element Results}

\begin{figure}[!htb]
\centering
  \includegraphics[width=0.90\linewidth]{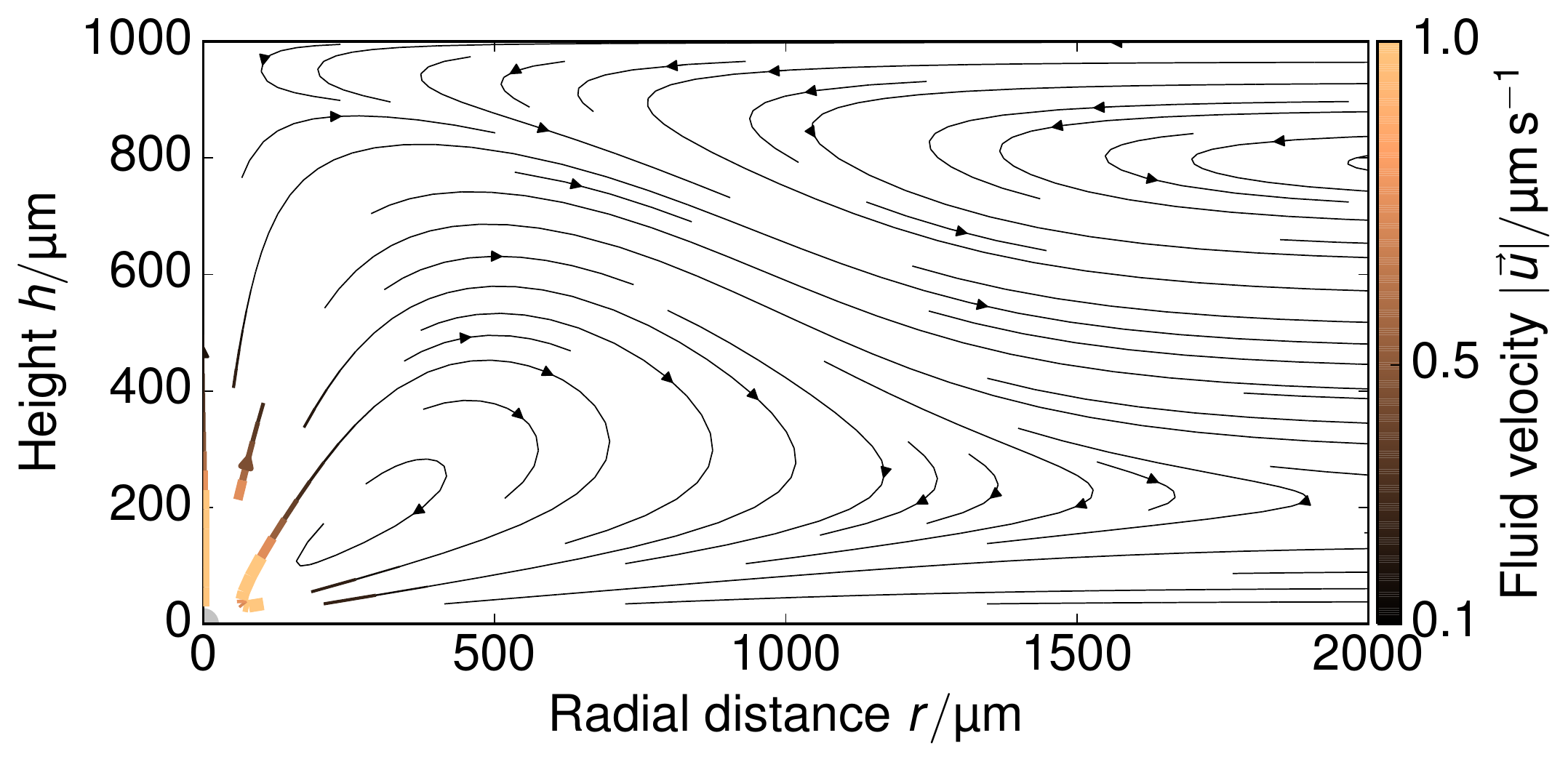}
  \caption{\label{fig:trajectories}Visualization of the FEM-calculated fluid flow in a $H=\SI{1}{\milli\meter}$ sample cell over a radial range of $\SI{2}{\milli\meter}$. The resin is shown in gray in the bottom-left corner, the direction of the fluid flow is shown with arrows, and color indicates the magnitude of the local velocity.}
\end{figure}

The FEM-computed fluid flow for the steady-state problem is shown in Fig.~\ref{fig:trajectories} for a large portion of the sample cell; we used $H = \SI{1}{\milli\meter}$. Both on the top and bottom wall the fluid flow is radially inward, due to the electroosmotic driving near the walls, with swirl-like patterns forming in the middle of the cell, due to the incompressibility of the fluid. 

\begin{figure}[!htb]
\centering
  \includegraphics[width=0.90\linewidth]{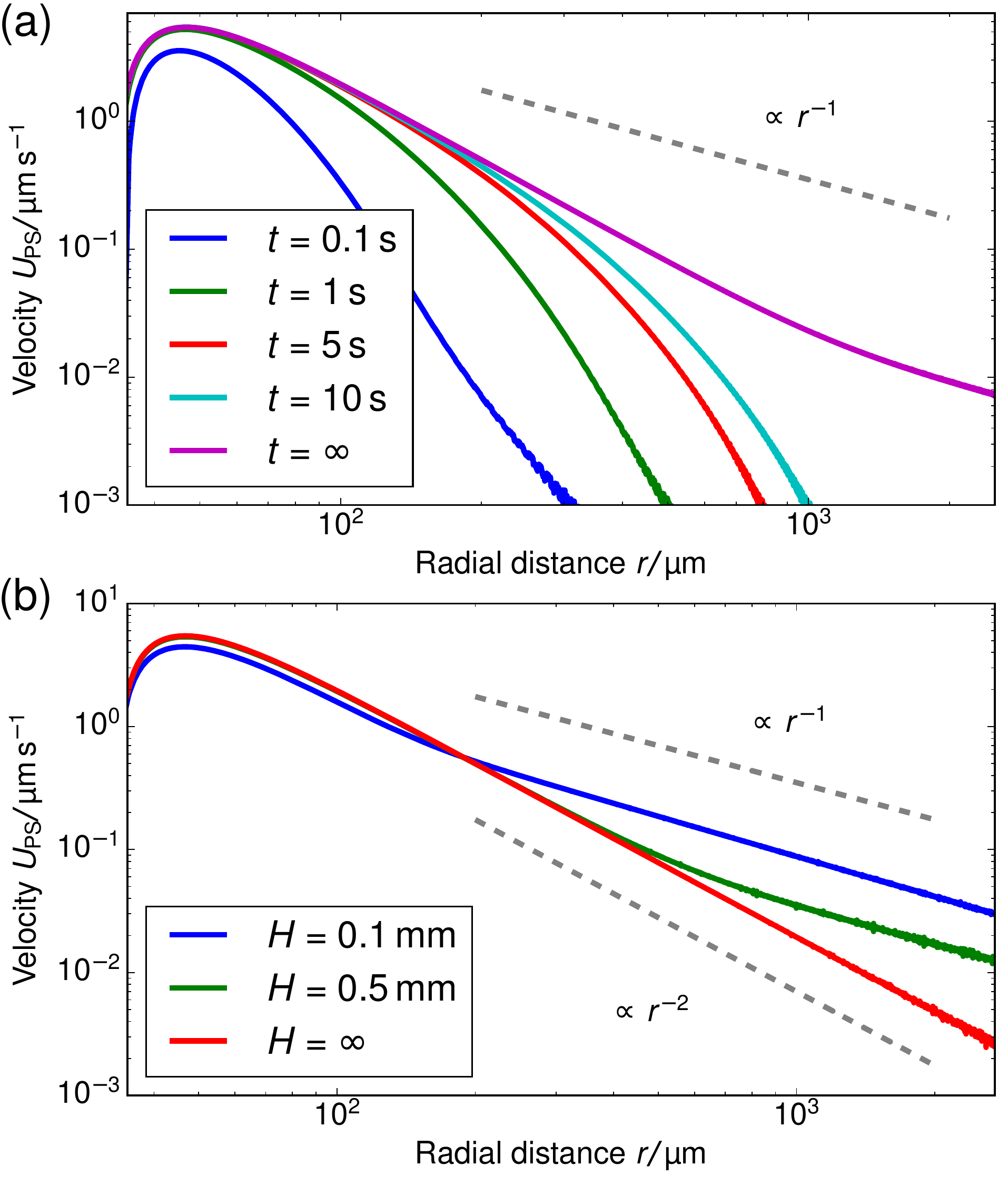}
  \caption{\label{fig:veloFEM}The tracer speed $U_{\mathrm{PS}}$ obtained using our numerical calculations for several configurations of the system. (a) The time-dependent solution for $U_{\mathrm{PS}}$ as a function of the radial distance $r$ for several times $t$, for PS7 tracers in a sample cell with height $H = \SI{0.2}{\milli\meter}$. The steady-state solution is given by the purple curve ($t = \infty$). The power-law decay sets in only for long times, see the far-right of the image. (b) The steady-state solution for several cell heights. The gray dashed lines serve as guides to the eye for the power-law decay.}
\end{figure}

Fitting the near-field tracer speed $U_{\mathrm{PS}}$ for a cation concentration of $\rho = \SI{1}{\micro\mole\per\liter}$ to the $H = \SI{1}{\milli\meter}$ profile in Fig.~\ref{fig:velo}, we found that $k_{\mathrm{ex}} = \SI{3.08E-6}{\meter\per\second}$ is sufficient to match the experimentally observed near-field speed --- compare Figs.~\ref{fig:velo} and~\ref{fig:veloFEM}a. We used this parameter throughout our simulations. This gives rise to an average surface flux of $j_{s} \approx \SI{7e-8}{\mole\square\per\meter\per\second}$, which corresponds closely to our back-of-the-envelope estimate in Section~\ref{sec:consider}. This shows that the experimentally observed tracer speeds can indeed be explained by ion exchange of trace amounts of cationic impurities in the $\si{\micro\mole\per\liter}$ range. 

First, we verified that our steady-state solution for the quasi-2D domain gives a reasonable result, when compared to the time-dependent simulations. We considered a cell height of $H = \SI{0.2}{\milli\meter}$ for this problem. This choice allowed us to reduce the number of mesh elements required compared to the typical experimental height $H = \SI{1.0}{\milli\meter}$ and thereby improve the computational time sufficiently to access second time scales. Figure~\ref{fig:veloFEM}a shows several time-dependent tracer speed curves, up to the maximum time of $t = \SI{10}{\second}$ that we could access with our FEM calculations (several days of computer run time). Note that for these times, the long-time, far-field power-law decay has not yet set in. We estimate the time for this decay to set in, using the time it takes \ce{H+} ions to diffuse a distance $H$:  $t_{\ce{H+}} = H^2/D_{\ce{H+}} \approx \SI{400}{\second}$. This time is short on the time scale of the experiment, but too long to access \textit{via} FEM calculations, which is why we consider analytic theory in Section~\ref{sec:theory}. Nevertheless, the near-field solution has begun to converge to the steady-state after \SI{10}{\second}. Considering the relatively short time scales compared to the length of the experiment, on which convergence should take place, we are justified in neglecting the time-dependence in the FEM calculations. 

Second, we considered the far-field $r^{-1}$ scaling in our steady-state simulations in Fig.~\ref{fig:veloFEM}b. For increasing $H$ there is an increasingly large intermediate range of 3D decay with $r^{-2}$. This is to be expected, because the minimum length (equivalently the time, in the time-dependent problem) that the ions travel before the quasi-2D decay sets in will increase with the height. This also explains the experimental observation of $r^{-2}$ scaling for the $H = \SI{10}{\milli\meter}$ sample cell in Fig.~\ref{fig:velo}, since for this height the transition time can be estimated to be $t_{\ce{H+}} = H^2/D_{\ce{H+}} \approx \SI{1.0E4}{\second}$, which is long compared to the experimental time scale (typical velocity measurements take place within an hour after sample preparation).

When solving the steady-state system of equations on the geometry of Fig.~\ref{fig:fem_setup}a, the flow field always decays with $r^{-1}$ in the far field. This is because in the steady-state problem a sufficient amount of time has passed for the ions to ``become aware'' of their confinement,~\textit{i.e.}, stationarity is analogous to $t \rightarrow \infty$ in the time-dependent problem. Therefore, the experimentally obtained transition from quasi-2D to 3D decay by increasing the height $H$, cannot be observed in such a simulation. To observe a $r^{-2}$ decay in the steady-state far field, an unbound domain must be simulated instead. Hence the need for the geometry of Fig.~\ref{fig:fem_setup}b.

Finally, comparing Figs.~\ref{fig:velo} and~\ref{fig:veloFEM}b, we find that the regime in which the power-law decay sets in is much closer to the resin in the experiment,~\textit{i.e.}, around $\SI{75}{\micro\meter}$. This is counterintuitive, since on the basis of simple geometry arguments one would expect the ions to become aware of their 2D confinement when the distance they have diffused becomes comparable to the confining height. This suggests that there are mechanisms by which the ions are transported faster than through diffusion alone. A clear candidate is advection via the fluid flow, as the flow field around the resin, see Fig.~\ref{fig:trajectories}, causes significant vertical displacement of the near-resin ions, provided the P{\'e}clet number is sufficiently large. We already estimated in Section~\ref{sec:electrokinetic} that this is likely the case. Unfortunately, the effect of advection cannot be incorporated in this work due to current limitations in computational performance for our FEM modelling, so that a quantitative match between theory and experiment is left for future study. Nevertheless, the qualitative behavior that we do capture together with our geometric arguments already provide important insights into the pumping mechanism.

\section{\label{sec:theory}Analytical Calculations}

In this section, we present an approximate, analytical solution to the electrokinetic equations on the domain of the sample cell. This allows us to obtain the time dependence and radial scaling of the flow in the far field. Our solution makes use of the equations provided in Section~\ref{sec:electrokinetic}, where we already made the following simplifying assumptions: (i) Advection can be neglected. (ii) The only ions present are $\ce{H+}$, $\ce{K+}$, and $\ce{Cl-}$. (iii) The diffusivities of \ce{K+} and \ce{Cl-} ions are equal. In order to make progress analytically, we also require: (iv) The perturbation of the ionic concentrations from their equilibrium distributions due to ionic fluxes generated by the resin bead are small compared to the background ionic concentrations --- this is likely to be strictly valid only at short times or far from the central bead. (v) Far enough from the resin bead, the solution can be treated as 2D, and the bead as a point ($\delta$-function) source. This means that our geometry is essentially a 2D disk, rather than the 3D cylinder segment of the FEM setup, with the fluxes of the species independent of the vertical position in the sample cell. (vi) The electrostatic Debye screening length $\kappa$ is much shorter than the relevant length scales of the problem, which are of order $H$. (vii) The resin bead produces constant, equal and opposite fluxes of {\ce{H+}} outwards and {\ce{K+}} inward.

Under the above assumptions, we can combine Eqs.~\eqref{eq:flux} and~\eqref{eq:NP} to obtain
\begin{align}
  \label{time linear 1} \partial_{t} x_{\ce{H+}}  &= D_{\ce{H+}} \nabla_{\mathrm{2D}}^{2} \left(x_{\ce{H+}}  + \psi\right) + \frac{\Gamma}{H\rho_{\ce{H+}}^{\infty}}\delta_{\mathrm{2D}} (\vec{r}) ;  \\
  \label{time linear 2} \partial_{t} x_{\ce{K+}}  &= D_{\ce{K+}} \nabla_{\mathrm{2D}}^{2} \left(x_{\ce{K+}}  + \psi\right) - \frac{\Gamma}{H\rho_{\ce{K+}}^{\infty}}\delta_{\mathrm{2D}} (\vec{r}) ; \\
  \label{time linear 3} \partial_{t} x_{\ce{Cl-}} &= D_{\ce{Cl-}}\nabla_{\mathrm{2D}}^{2} \left(x_{\ce{Cl-}} - \psi\right) ,
\end{align}
where we define dimensionless concentrations ${x_{i} \equiv (\rho_{i}-\rho_{i}^{\infty})/\rho_{i}^{\infty}}$, the dimensionless electrostatic potential $\psi=\Phi e/(k_{\mathrm{B}}T)$, and $\nabla_{\mathrm{2D}}^{2}$ is the 2D Laplacian. From the linear approximation (iv), we have kept only terms up to linear order in $x_{i}$ and $\psi$, and from the no-advection approximation (i), we have neglected the term in $\vec{u}$. The final term in Eqs.~\eqref{time linear 1} and~\eqref{time linear 2} represents the steady production of $\ce{H+}$ and consumption of $\ce{K+}$ at the origin. Here, the 2D radial vector $\vec{r}=\vec{x}+\vec{y}$ ($r = \vert \vec{r} \vert$) and $\delta_{\mathrm{2D}}$ is the 2D $\delta$-function, which is normalized so that $\iint \delta(\vec{r})\mathrm{d}^2\vec{r}=1$, with the integral running over the whole plane. $\Gamma$ is the total production rate of $\ce{H+}$ in molecules$\;$\si{\per\second}. Note that we do not make $\Gamma$ dependent on $\rho_{\ce{K+}}$, as in the FEM model, to avoid complicating our calculation.

Linearizing the Poisson equation, Eq.~\eqref{eq:Poisson} and defining the dimensionless background concentrations ${\alpha_{i} \equiv \rho_{i}^{\infty} /(\sum_{j} \rho_{j}^{\infty} z_{j}^{2})}$ gives
\begin{align}
  \label{rho linear} \kappa^{2} \sum_{i}\alpha_{i} z_{i} x_{i}= -\nabla_{\mathrm{2D}}^{2} \psi , 
\end{align}
where the inverse Debye length $\kappa$ is given by Eq.~\eqref{eq:Debye}.

We now apply the thin-Debye-layer approximation (vi). Since we are interested in distances from the origin $r \gg H$, the capillary height, this approximation can be quantified as $\kappa r\gg 1$, and the approximation involves making an expansion to lowest order in the small parameter $1/(\kappa r)$. Now, from Eqs.~\eqref{time linear 1}-\eqref{time linear 3}, we must have that $\psi$ and $x_{i}$ are of similar order. However, if we multiply Eq.~\eqref{rho linear} by $r^{2}$, we see that the right-hand side is of order unity since $\nabla_{\mathrm{2D}}^{2} = \mathcal{O}(r^{-2})$, but the left-hand side is of order $\kappa^2 r^2$. This means that, for consistency, the sum on the left-hand side of Eq.~\eqref{rho linear} must be zero to lowest order in $1/(\kappa r)$,~\textit{i.e.},
\begin{align}
  \label{charge balance 1} \sum_{i} \alpha_{i} z_{i} x_{i} &= 0 + \mathcal{O}\left( (\kappa r)^{-2} \right). 
\end{align}
That is, the charge density is approximately zero everywhere outside a thin Debye layer close to the capillary surface. Note that this does not mean that $\nabla_{\mathrm{2D}}^{2} \psi = 0$, as from Eq.~\eqref{rho linear} it follows that the leading order term in $\nabla_{\mathrm{2D}}^{2} \psi$ is equal to the finite next-to-leading-order term in the charge density. For simplicity, we write $\alpha_{\ce{H+}} = \alpha$ and, from the condition of charge balance in the background concentrations, $\alpha_{\ce{K+}} = 1/2 - \alpha$ and $\alpha_{\ce{Cl-}} = 1/2$. Then Eq.~\eqref{charge balance 1} can be rewritten as
\begin{align}
  \label{charge balance} 2 x_{\ce{H+}} \alpha + x_{\ce{K+}}(1 - 2\alpha ) &= x_{\ce{Cl-}} . 
\end{align}

Equations~\eqref{time linear 1}-\eqref{time linear 3} together with Eq.~\eqref{charge balance} represent a linear system of equations that we will now solve. From linear combinations of Eqs.~\eqref{time linear 1}-\eqref{time linear 3}, we can eliminate $\psi$, and Eq.~\eqref{charge balance} allows us to also eliminate $x_{\ce{Cl-}}$. This leaves us with two equations for $x_{\ce{H+}}$ and $x_{\ce{K+}}$
\begin{widetext}
\begin{align}
  \left( \frac{1}{D_{\ce{H+}}}+\frac{2\alpha}{D_{\ce{Cl-}}} \right)\partial_{t} x_{\ce{H+}} + \left(\frac{1-2\alpha}{D_{\ce{Cl-}}}\right)\partial_{t} x_{\ce{K+}} &= \nabla_{\mathrm{2D}}^{2} \left[ (1+2\alpha)x_{\ce{H+}} + (1-2\alpha)x_{\ce{K+}} \right] + \frac{\Gamma}{H\rho_{\ce{H+}}^{\infty} D_{\ce{H+}}}\delta_{\mathrm{2D}}(\vec{r}); \\
  \left( \frac{2\alpha }{D_{\ce{Cl-}}} \right)\partial_{t} x_{\ce{H+}} + \left(\frac{1}{D_{\ce{K+}}}+\frac{1-2\alpha}{D_{\ce{Cl-}}}\right)\partial_{t} x_{\ce{K+}} &= \nabla_{\mathrm{2D}}^{2} \left[ 2\alpha x_{\ce{H+}} + 2(1-\alpha)x_{\ce{K+}} \right] - \frac{\Gamma}{H\rho_{\ce{K+}}^{\infty} D_{\ce{K+}}}\delta_{\mathrm{2D}}(\vec{r}).
\end{align}
\end{widetext}
We solve these equations using the ansatz functions
\begin{align}
  \label{final solution 1} x_{\ce{H+}} &= \frac{\Gamma}{\rho_{\ce{H+}}^{{\infty}} H} \left[ A f_{1}(r,t) + B f_{2}(r,t) \right]; \\
  \label{final solution 2} x_{\ce{K+}} &= \frac{\Gamma}{\rho_{\ce{K+}}^{{\infty}} H} \left[ C f_{1}(r,t) + E f_{2}(r,t) \right], 
\end{align}
where $A$, $B$, $C$, and $E$ are constants, and $f_{1}(r,t)$ and $f_{2}(r,t)$ are given by the time-integral of the Green's function for the 2D diffusion equation~\cite{ortiz16,shin16}, which accounts for the constant point-source at the origin
\begin{align}
  f_{m} &= \int_{0}^{t} \frac{1}{4 \pi \tilde{D}_{m} t'} \exp \left( \frac{-r^{2}}{4 \tilde{D}_{m} t'} \right)\mathrm{d}t' ,
\end{align}
where the $\tilde{D}_{m}$, $m \in \{1,2\}$ are effective diffusivities to be determined. Directly solving for these effective diffusivities in general yields extremely unwieldy expressions, so instead we first make the simplifying assumption (iii) that $D_{\ce{K+}} = D_{\ce{Cl-}}$. This gives
\begin{align}
  \tilde{D}_{1} &= D_{\ce{K+}}; \\
  \tilde{D}_{2} &= \frac{ D_{\ce{H+}}D_{\ce{K+}} }{ D_{\ce{K+}} + \alpha (D_{\ce{H+}} - D_{\ce{K+}}) }.
\end{align}
We then solve for the constant terms in Eq.~\eqref{final solution 2}, yielding
\begin{align}
  A &= 0; \\
  B &= 1; \\
  C &= \frac{1}{2(1-\alpha )}; \\
  E &= \frac{1-2\alpha}{2(1-\alpha)}.
\end{align}
Plugging these constants into Eqs.~\eqref{final solution 1}-\eqref{final solution 2} and using ${x_{i} \equiv (\rho_{i}-\rho_{i}^{\infty})/\rho_{i}^{\infty}}$ allows us to obtain the time-dependent ion density profiles. To calculate $\psi$, we take the time derivative of Eq.~\eqref{charge balance} and use this to eliminate the left-hand side from Eqs.~\eqref{time linear 1}-\eqref{time linear 3} to obtain, after some algebra
\begin{align}
  \psi &= -\frac{ (D_{\ce{H+}} - D_{\ce{K+}})\alpha }{ (D_{\ce{H+}}-D_{\ce{K+}})\alpha + D_{\ce{K+}} } x_{\ce{H+}} .
\end{align}
Using $\Phi = k_{\mathrm{B}} T \psi/e$ this yields
\begin{align}
  \label{phi} \Phi &= -\left( \frac{ D_{\ce{H+}}-D_{\ce{K+}} }{ D_{\ce{K+}}D_{\ce{H+}} } \right) \frac{e\Gamma}{4 \pi \epsilon_{r} \epsilon_{0} H \kappa^{2}}\int_0^t \frac{1}{t'}\exp\left(\frac{-r^{2}}{4 \tilde{D}_{2} t'}\right)\mathrm{d}t' . 
\end{align}

The potential in Eq.~\eqref{phi} will generate an equal slip-velocity $\vec{u}_{\mathrm{slip}} = (\zeta\epsilon/\eta)\vec{\nabla}_{\mathrm{2D}}\Phi$ on both the upper and lower surfaces of the channel; we are sufficiently far away from the resin that the asymmetry caused by it being glued to the bottom wall should not strongly affect the flow field. From Eq.~\eqref{phi} we then obtain
\begin{align}
 \label{eq:slip} \vec{u}_{\mathrm{slip}} &= -\frac{1}{r} \left[ \left( \frac{ D_{\ce{H+}}-D_{\ce{K+}} }{ D_{\ce{K+}}D_{\ce{H+}} } \right) \frac{\zeta e\Gamma}{2\pi\kappa^{2}\eta H} \exp\left( \frac{-r^{2}}{4 \tilde{D}_{2} t} \right) \hat{r} \right] . 
\end{align}
Strictly speaking, $\vec{u}_{\mathrm{slip}}$ is the velocity at the outer edge of the Debye layer. However, in the thin Debye limit, we can take $\vec{u}_{\mathrm{slip}}$ to be the fluid velocity on the wall itself. In the bulk of the channel, the fluid flow will vary with a typical length scale $H$, the capillary height. Therefore, we can use $\vec{u}_{\mathrm{slip}}$ as a good estimate for the velocity $\vec{u}_{\mathrm{PS}}(z)$ of a tracer particle located a small distance $z \ll H$ above the capillary surface, with a fractional error $\mathcal{O}(z/H)<10^{-2}$ for the channels used here.

\begin{figure}[!htb]
\centering
  \includegraphics[width=0.90\linewidth]{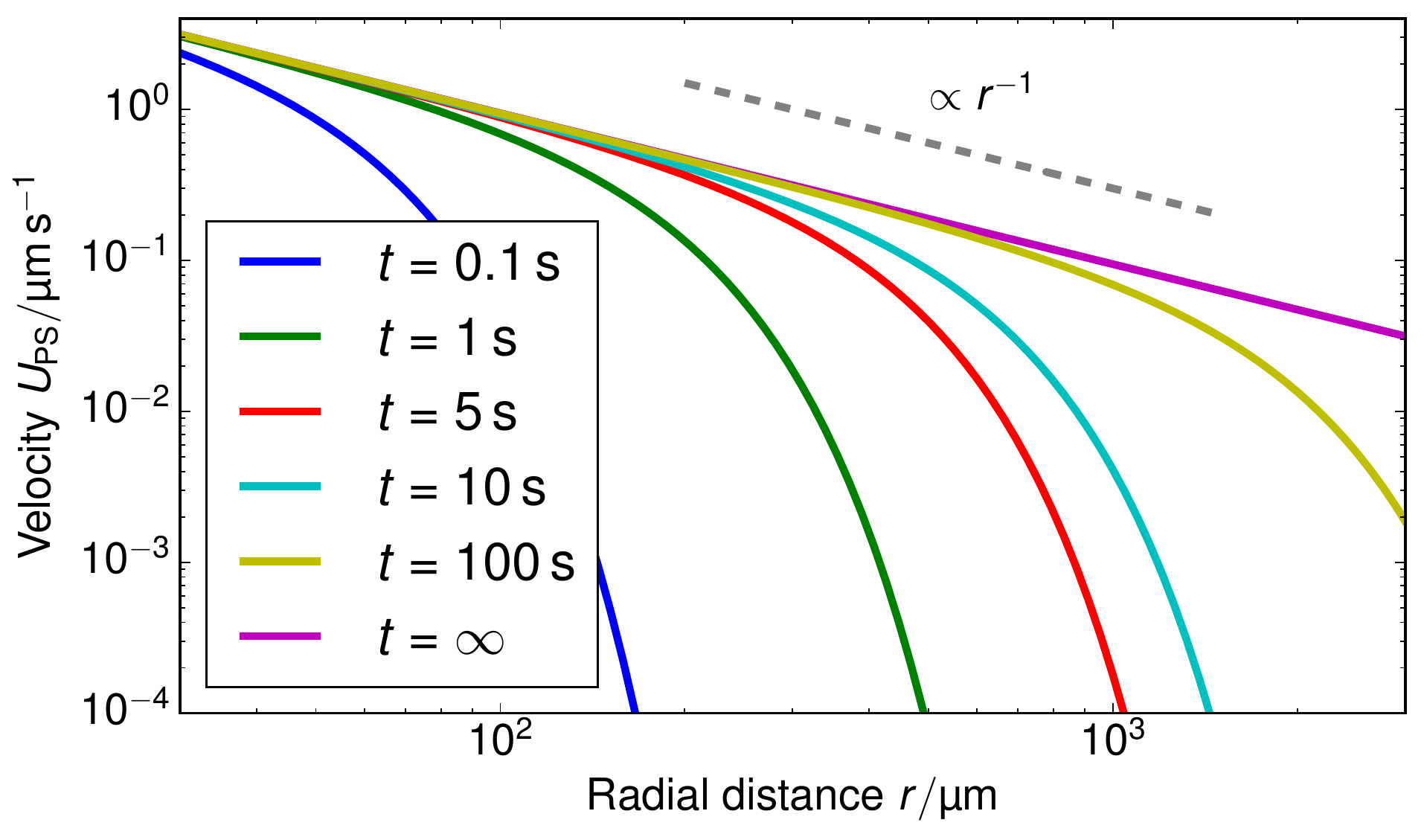}
  \caption{\label{fig:veloanalytic}The tracer speed $U_{\mathrm{PS}}$ obtained using our analytic theory as a function of the radial distance $r$ for several times $t$, for PS7 tracers in a sample cell with height $H = \SI{0.2}{\milli\meter}$, $\mathrm{pH} = 7$, and $\Gamma = \SI{2e8}{\per\second}$; this models the setup of Fig.~\ref{fig:veloFEM}a. The steady-state solution is given by the purple curve ($t = \infty$). The gray dashed line serves as a guide to the eye for the power-law decay.}
\end{figure}

Figure~\ref{fig:veloanalytic} shows the tracer speed $U_{\mathrm{PS}}$ as a function of the distance as calculated using Eq.~\eqref{eq:slip} for a sample cell with $H = \SI{0.2}{\milli\meter}$ and other parameters the same as in Fig.~\ref{fig:veloFEM}a. Since the theory does not use the same expression for the surface fluxes (vii), we selected a molecular exchange rate of $\Gamma = \SI{2e8}{\per\second}$ to give reasonable speed correspondence when compared to the results in Fig.~\ref{fig:veloFEM}a. At long times,~\textit{i.e.}, for $t \gg r^{2}/\tilde{D}_{2}$, the exponential term approaches unity, so we obtain the $1/r$ scaling observed experimentally. At short times, the exponential term dominates, so we obtain the rapid radial decays seen in the FEM calculations in Fig.~\ref{fig:veloFEM}a. 

Note that, as we would expect for a 2D system, the electrostatic field in Eq.~\eqref{phi} (as well as the ion density profiles) is not in steady state --- the integral approaches $\log(t)$ in the limit of large $t$. Nevertheless, the flow field in Eq.~\eqref{eq:slip} does approach a steady-state solution, see Fig.~\ref{fig:veloanalytic}, because it scales with the gradient $\vec{\nabla}_{\mathrm{2D}}\phi$, which is in steady state. This is in line with our experimental observations, where we observed the same tracer-velocity trends after \SI{24}{\hour} of pumping, albeit with decrease in speed by a factor of two. The latter can be attributed to depletion of the trace amounts of cations in the cell or a reduction in the effectiveness with which the resin exchanges ions. These results further underpin our conclusion that the microfluidic pumping is driven by ion exchange of trace amounts of cationic species in the sample cell.

The correspondence to the FEM calculations (Fig.~\ref{fig:veloFEM}a) is semi-quantitative, despite the additional simplifying assumptions. However, it should be noted that we cannot make a prediction for the near field using the theory, due to the quasi-2D assumption (v), nor do we account for advection of the solutes (i). This means that we cannot expect, nor do we find, quantitative agreement with the experiment.

\section{\label{sec:concl}Discussion and Outlook}

In summary, we have introduced and characterized an ion-exchange-resin-based microfluidic pump. The striking feature of this pump is that it operates in \si{\micro\mole\per\liter} ionic concentrations for periods exceeding \SI{24}{\hour} and yet manages to produce fluid flows with speeds of several \si{\micro\meter\per\second} over hundreds of \si{\micro\meter}, without strongly modifying its environment.

We demonstrated that our pump uses trace amounts of cations to generate fluid flow using a combination of tracer velocimetry experiments, analytic electrokinetic theory, and finite-element-method simulations. These together show that fluid flow is achieved via electroosmosis, by the exchange of cations for protons in its interior. The difference in ionic mobility between the cations and the protons, for which they are exchanged, sets up a diffusion potential that points towards the resin and causes fluid flow in this direction. The speed of pumping can be modified by varying the bulk cation concentration in the \si{\micro\mole\per\liter} range. 

Our pump has several advantages over other microfluidic pumps that also exploit diffusion-potential-based electroosmosis,~\textit{e.g.}, small pieces of salt that slowly dissolve~\cite{ibele09, mcdermott12}. Firstly, ion-exchange pumps only modify the nature of the bulk ions, not their net concentration. Secondly, as a pump dissolves, it might change shape, thereby inducing an undesirable directionality to the pumping. Our system does not have this disadvantage, as our spherical ion-exchange resins retain their shape throughout. Thirdly, ion exchange using protons as the exchangeable cation has the advantage of setting up significant diffusion potentials (and hence flow), due to the strong difference in diffusivity between the proton and any exchanged cationic species. Finally, the ion-exchange-resin pump functions for very long times in a low-ionicity medium --- over \SI{24}{\hour} --- compared to the much shorter operating times of dissolving micropumps, which were indicated to be around \SI{20}{\minute} in Ref.~\cite{mcdermott12}.

The range of our pump can be tuned via the height of the sample cell to give rise to either 3D or quasi-2D decay of the far-field flow velocity,~\textit{i.e.}, power-law decay with the relevant exponent. We have thus demonstrated that significant microfluidic pumping can be achieved at very low fuel (cation) concentrations and can be sensitively tuned via the geometry. This tunability can be exploited to self-assemble single colloidal crystals~\cite{niu16b}. 

In our modeling of the experiment it proved necessary to ignore the advective contributions to the ionic fluxes. This simplifying assumption is the likely cause of the quantitative (but not qualitative) differences between the theory and experiment.  We argue in favor of including advective (fluid flow) contributions to the ionic fluxes in any future modeling of these systems, as this is important in understanding the near-field fluid flow around the ion-exchange resin, and to extract the kinetics of ion exchange from far-field flow and concentration profile measurements.

Presently, we are only able to indicate that the ion-exchange process is likely diffusion limited in our system. Future experimental work will focus on pH measurements to quantify the exchange process, while the nature of the decay in these concentration profiles will be further examined using numerical approaches. For the latter, the use of a boundary-layer approach to rescale the high P{\'e}clet number regime and make these problems computationally tractable will be explored. Furthermore, capturing the near-field flow accurately will be relevant to understanding the formation of self-assembled cooperative swimmers based on mobile ion-exchange resins and tracer particles~\cite{niu16a}.

In conclusion, our system showcases the significance of very small ionic concentrations and fluxes in microfluidic settings. This suggests that such fluxes may be responsible for flow and motion in a much wider range of out-of-equilibrium systems, such as for chemical swimmers and in biological processes, and should be considered in future modeling thereof.

\textit{Acknowledgements} --- JdG gratefully acknowledges financial support from a Marie Sk{\l}odowska-Curie Intra European Fellowship (G.A. No. 654916) within Horizon 2020. JdG and CH, as well as RN, DB, and TP, further thank the DFG for funding through the SPP 1726 ``Microswimmers --- From Single Particle Motion to Collective Behavior'' (HO1108/24-1 \& Pa459/18-1).

\end{document}